\documentclass[11pt, reqno]{amsart}
\usepackage{amsfonts}
\usepackage{amssymb}
\usepackage{latexsym}
\usepackage[final]{hyperref}
\usepackage[]{enumerate}
\usepackage{verbatim}
\usepackage[english]{babel} 
\usepackage{blindtext}
\usepackage[nohyperlinks,nolist]{acronym}
\usepackage{listings}
\usepackage{multirow}

\usepackage[table]{xcolor}%

\usepackage{graphicx}       
\usepackage[utf8]{inputenc} 
\usepackage{amsmath}

\usepackage{tikz}
\usepackage{cmap}
\usepackage{ulem}
\usepackage{soul}
\usepackage{comment}
\usepackage{fancybox}
\usepackage{relsize}
\usepackage{pifont}
\usepackage{subcaption}
\usepackage{enumitem}

\newcommand{\Abs}[1]{\left\vert #1 \right\vert}
\newcommand{\abs}[1]{\vert #1 \vert}

\newcommand{\norm}[1]{\left\Vert #1 \right\Vert}

\DeclareMathOperator{\erf}{erf}

\theoremstyle{definition}

\theoremstyle{remark}

\begin{acronym}
\acro{AO}{atomic orbital}
\acro{API}{Application Programmer Interface}
\acro{AUS}{Advanced User Support}
\acro{BEM}{Boundary Element Method}
\acro{BO}{Born-Oppenheimer}  
\acro{CBS}{complete basis set}
\acro{CC}{Coupled Cluster}
\acro{CTCC}{Centre for Theoretical and Computational Chemistry}
\acro{CoE}{Centre of Excellence}
\acro{DC}{dielectric continuum}  
\acro{DCHF}{Dirac-Coulomb Hartree-Fock}  
\acro{DFT}{density functional theory}  
\acro{DKH}{Douglas-Kroll-Hess}
\acro{EFP}{effective fragment potential}
\acro{ECP}{effective core potential}
\acro{EU}{European Union}
\acro{FFT}{Fast Fourier Transform}
\acro{GGA}{generalized gradient approximation}
\acro{GPE}{Generalized Poisson Equation}
\acro{GTO}{Gaussian Type Orbital}
\acro{HF}{Hartree-Fock}  
\acro{HPC}{high-performance computing}
\acro{Hylleraas}[HC]{Hylleraas Centre for Quantum Molecular Sciences}
\acro{IEF}{Integral Equation Formalism}
\acro{IGLO}{individual gauge for localized orbitals}
\acro{KB}{kinetic balance}
\acro{KS}{Kohn-Sham}
\acro{LAO}{London atomic orbital}
\acro{LAPW}{linearized augmented plane wave}
\acro{LDA}{local density approximation}
\acro{MAD}{mean absolute deviation}
\acro{maxAD}{maximum absolute deviation}
\acro{MM}{molecular mechanics}  
\acro{MCSCF}{multiconfiguration self consistent field}
\acro{MPA}{multiphoton absorption}
\acro{MRA}{multiresolution analysis}
\acro{MSDD}{Minnesota Solvent Descriptor Database}
\acro{MW}{multiwavelet}
\acro{NAO}{numerical atomic orbital}
\acro{NeIC}{nordic e-infrastructure collaboration}
\acro{KAIN}{Krylov-accelerated inexact Newton}
\acro{NMR}{nuclear magnetic resonance}
\acro{NP}{nanoparticle}  
\acro{NS}{non-standard}  
\acro{OLED}{organic light emitting diode}
\acro{PAW}{projector augmented wave}
\acro{PBC}{Periodic Boundary Condition}
\acro{PCM}{polarizable continuum model}
\acro{PW}{plane wave}
\acro{QC}{quantum chemistry}  
\acro{QM/MM}{quantum mechanics/molecular mechanics}  
\acro{QM}{quantum mechanics}  
\acro{RCN}{Research Council of Norway}
\acro{RMSD}{root mean square deviation}
\acro{RKB}{restricted kinetic balance}
\acro{SC}{semiconductor}
\acro{SCF}{self-consistent field}
\acro{STSM}{short-term scientific mission}
\acro{SAPT}{symmetry-adapted perturbation theory}
\acro{SERS}{surface-enhanced raman scattering}
\acro{WPREL}[WP1]{Work Package 1}
\acro{WPROP}[WP2]{Work Package 2}
\acro{WPAPP}[WP3]{Work Package 3}
\acro{WP}{Work Package}  
\acro{X2C}{exact two-component}
\acro{ZORA}{zero-order relativistic approximation}
\acro{ae}{almost everywhere}
\acro{BVP}{boundary value problem}
\acro{PDE}{partial differential equation}
\acro{RDM}{1-body reduced density matrix}
\acro{SCRF}{self-consistent reaction field}
\acro{IEFPCM}{Integral Equation Formalism \ac{PCM}}
\acro{FMM}{fast multipole method}
\acro{DD}{domain decomposition}
\acro{TRS}{time-reversal symmetry}
\acro{SI}{Supporting Information}
\acro{DHF}{Dirac--Hartree--Fock}
\acro{MO}{molecular orbital}
\end{acronym}

\title
[
    Heat semigroup representation of Laplacian
]
{
    Heat semigroup representation of Laplacian
}

\author[Dinvay]{Evgueni Dinvay}

\email{ evgueni.dinvay@uit.no }

\address
{
    Department of Chemistry
    \\
    UiT The Arctic University of Norway
    \\
    PO Box 6050 Langnes
    \\
    N-9037 Tromsø
    \\
    Norway
}

\date{January 15, 2025}

\begin{document}

\begin{abstract}
This work introduces novel numerical algorithms for computational quantum mechanics,
grounded in a representation of the Laplace operator
--
frequently used to model kinetic energy in quantum systems
--
via the heat semigroup.
The key advantage of this approach lies in its computational efficiency,
particularly within the framework of multiresolution analysis,
where the heat equation can be solved rapidly.
This results in an effective and scalable method for evaluating kinetic energy in quantum systems.
The proposed multiwavelet-based Laplacian approximation is tested
through two fundamental quantum chemistry applications:
self-consistent field calculations and the time evolution of Schrödinger-type equations. 
\end{abstract}

\maketitle
\section{Introduction}
\setcounter{equation}{0}

Consideration is given to two  of the most fundamental Schr\"odinger-type equations
\begin{equation}
\label{time_schrodinger}
    i \partial_t \psi
    =
    - \frac 12 \Delta \psi
    + V( \psi, x, t)
\end{equation}
and the corresponding eigenvalue problem
\begin{equation}
\label{ground_schrodinger}
    - \frac 12 \Delta \psi
    + V( \psi, x)
    =
    E \psi
    ,
\end{equation}
which underlie many models in quantum physics.
A complex-valued wavefunction $\psi$, solving either \eqref{time_schrodinger} or \eqref{ground_schrodinger},
depends on the space variable $x \in \mathbb R^d$ and,
in the case of \eqref{time_schrodinger}, on the time variable $t > 0$.
The most natural dimension for physical systems is $d = 3$.
Each of these equations can be treated independently or as part of a system of equations,
coupled through the functional $V$, as in many-electron theory for atoms and molecules.

The primary focus of the current work is to develop an effective numerical approximation
for the Laplace operator $\Delta = \partial_{x_1}^2 + \ldots + \partial_{x_d}^2$.
It is well known that, when applied to a sufficiently smooth function $f(x)$,
the heat semigroup $e^{t \Delta}$,
that is the fundamental solution of the heat equation $\partial_t u = \Delta u$,
can be approximated by
\(
    e^{t \Delta} f \approx f + t \Delta f
\)
for small $t$.
In other words, $\Delta$ can be regarded as a linear combination of the identity operator
and a convolution $e^{t \Delta}$ with a sharply peaked Gaussian kernel
\(
    (4 \pi t)^{- d/2} \exp( - |x|^2 / (4t))
    .
\)
One can obtain a more precise expansion
\begin{equation}
\label{Delta_f_expansion}
    \Delta f
    =
    \frac 1{2 t}
    \left(
        4 e^{t \Delta} - e^{2 t \Delta} - 3
    \right)
    f
    +
    O \left( t^2 \right)
\end{equation}
as $t \to 0$.
Below we elaborate more on different representations to increase the order of
the Laplace operator approximation.

Normally, transforming the evaluation of second-order derivatives
into a convolution problem is not particularly advantageous.
However, there exist spatial discretization methods that are better optimized
for performing convolutions than for iterative applications of derivatives.
One such method is \ac{MRA} \cite{Alpert_Beylkin_Gines_Vozovoi2002},
which will be reviewed in the next section.
Here, we note that this adaptive discretization technique is particularly useful in computational chemistry,
where it effectively handles the singular Coulomb interaction between an electron and the nucleus,
as well as the resulting cusp in orbitals
\cite{
    Frediani_Fossgaard_Fla_Ruud,
    Harrison_Fann_Yanai_Gan_Beylkin2004,
    Jensen_Saha_elephant2017,
    Yanai_Fann_Gan_Harrison_Beylkin2004}.
Moreover, its success in chemistry applications is largely due to the approximation of the resolvent:
\begin{equation}
\label{resolvent_approximation}
    \left( - \Delta + \mu^2 \right)^{-1}
    \approx
    \sum_{j} c_j e^{t_j \Delta}
\end{equation}
based on the heat semigroup \cite{Harrison_Fann_Yanai_Beylkin2003}.
The expansion \eqref{resolvent_approximation} enables a natural treatment of the singularity
in the Green's function, the kernel of the resolvent,
while also facilitating the separation of variables $x_1, \ldots, x_d$.
Multiresolution quantum chemistry is rapidly evolving,
and we refer readers to a comprehensive review for a broader perspective \cite{Bischoff2019}.
An overview of the latest advancements will also appear in
an upcoming paper \cite{Tantardini_Dinvay_Pitteloud_Gerez_Jensen_review2024}.

As mentioned above,
the primary aim of this contribution is to provide an approximate heat semigroup representation,
similar to \eqref{resolvent_approximation},
for the second derivatives that constitute the Laplace operator.
This representation is developed in Section \ref{Laplacian_decomposition_section}.
It can be further discretized using a multiscale technique and applied in self-consistent field calculations,
as demonstrated in Section \ref{Self_consistent_field_section}.
It can be regarded as a direct way of using \ac{MRA} for the ground state search,
compared to the Green's function treatment \cite{Jensen_Durdek_Bjorgve_Wind_Fla_Frediani2023}
based on the Gaussian expansion \eqref{resolvent_approximation}
and by now well established in multiresolution chemistry calculations.
The new treatment of the ground state problem does not outperform the resolvent's approach,
and at the current stage may serve mostly as a proof of principle.
However,
the introduced here numerical representation of Laplacian allows to incorporate an alternative
way of doing time evolution simulations for quantum systems in the \ac{MRA} framework,
by expanding
the free-particle propagator
\(
    e^{i t \Delta}
\)
in the Taylor series.
Indeed, the two existing approaches \cite{Dinvay_Zabelina_Frediani2024, Vence_Harrison_Krstic2012}
are technically challenging and
need high order polynomials while multiscaling adaptively
the semigroup
\(
    e^{i t \Delta}
    .
\)
Moreover, the treatment of oscillatory integrals conducted in \cite{Dinvay_Zabelina_Frediani2024}
is restricted by the use of large time steps.
It turns out that the polynomial order restriction is irrelevant for
the considered here in Section \ref{Free_particle_propagator_section}
the heat semigroup representation of the exponential operator
\(
    e^{i t \Delta}
    .
\)
Whereas the time steps should be on the contrary small,
which complements significantly the results in \cite{Dinvay_Zabelina_Frediani2024}.

\section{Multiresolution analysis}
\label{Multiresolution_analysis_section}
\setcounter{equation}{0}

This section  is devoted to the multiscale approximation by piece wise polynomials,
a tool that can be particularly useful for fast evaluations of convolutions
with Gaussian kernels.
The main focus is on the one dimensional case $d = 1$ with the computational domain $[0, 1]$.
The function space $L^2(0, 1)$ is embedded into $L^2(\mathbb R)$
by the trivial extension outside the interval $(0, 1)$.
Let $\phi_0, \ldots, \phi_{\mathfrak k - 1}$ be an orthonormal set of polynomials in $L^2(0, 1)$.
Two principal examples used in multiwavelet calculations are
formed from either shifted Legendre polynomials or interpolation polynomials
associated to Gauss quadrature points \cite{Alpert_Beylkin_Gines_Vozovoi2002}.
These polynomials, referred to as scaling functions, span a $\mathfrak k$-dimensional subspace
$V_0^{\mathfrak k}$ in $L^2(\mathbb R)$.
We define a space $V_n^{\mathfrak k}$ spanned
by $2^n \mathfrak k$ functions which are obtained from the scaling functions by
dilation and translation
\begin{equation}
\label{scaling_dilation_translation}
    \phi_{jl}^n(x)
    =
    2^{n/2}
    \phi_j( 2^n x - l)
    , \quad j = 0, \ldots, \mathfrak k - 1
    , \quad l = 0, \ldots, 2^n - 1
    .
\end{equation}
In other words,
a function $f \in V_n^{\mathfrak k}$ provided that on each dyadic interval
\(
    (l / 2^n, (l + 1) / 2^n)
\)
with
\(
    l = 0, 1, \ldots, 2^n - 1
\)
it is a polynomial of degree less than $\mathfrak k$ and it is zero elsewhere.
The union
\(
    \bigcup_{n = 0}^{\infty} V_n^{\mathfrak k}
\)
is dense in $L^2(0, 1)$.
We refer to the given fixed number $\mathfrak k$ as the order of MRA and to
the integer variable $n$ as a scaling level.

The multiwavelet space $W_n^{\mathfrak k}$ is defined as
the orthogonal complement of $V_n^{\mathfrak k}$ in $V_{n + 1}^{\mathfrak k}$.
It is enough to choose a basis
$\psi_0, \ldots, \psi_{\mathfrak k - 1}$ in $W_0^{\mathfrak k}$
and then transform it in line with the general rule \eqref{scaling_dilation_translation},
in order to get an orthonormal basis in each $W_n^{\mathfrak k}$.
The primary example of multiwavelet functions was constructed in \cite{Alpert1993},
providing an additional vanishing moment property: specifically,
each $\psi_j(x)$ is orthogonal to polynomials of degree less than $j + \mathfrak k$.

The expansion coefficients of a general function $f$ in the scaling basis $\phi_{jl}^n$
of $V_n^{\mathfrak k}$ are the integrals
\begin{equation}
\label{scaling_coefficients}
    s_{jl}^n(f)
    =
    2^{-n/2}
    \int_0^1
    f( 2^{-n} (x + l) ) \phi_j(x)
    dx
    , \quad j = 0, \ldots, \mathfrak k - 1
    , \quad l = 0, \ldots, 2^n - 1
    ,
\end{equation}
that are evaluated numerically using the Gauss-Legendre quadrature.
The multiwavelet expansion coefficients $d_{jl}^n(f)$ of $f$ in the basis $\psi_{jl}^n$
of $W_n^{\mathfrak k}$ are obtained from the following relation
\begin{equation}
\label{decomposition_step}
    \begin{pmatrix}
        s_l^n
        \\
        d_l^n
    \end{pmatrix}
    =
    U
    \begin{pmatrix}
        s_{2l}^{n + 1}
        \\
        s_{2l + 1}^{n + 1}
    \end{pmatrix}
\end{equation}
with vectors $s_l^n$, $d_l^n$ consisting of coefficients 
$s_{jl}^n$, $d_{jl}^n$, $j = 0, \ldots, \mathfrak k - 1$, accordingly.
The unitary matrix $U$ is referred to as the forward wavelet transform \cite{Cohen_1992}.
It is uniquely defined by the choice of bases in $V_0^{\mathfrak k}$ and $W_0^{\mathfrak k}$.

It constitutes the multiwavelet representation for functions of a single variable.
For $d > 1$,
the scaling spaces $V_n^{\mathfrak k, d}$ are defined as the tensor products of
the one dimensional analogues $V_n^{\mathfrak k}$.
The wavelet space $W_n^{\mathfrak k, d}$ is then defined
as the orthogonal complement of $V_n^{\mathfrak k, d}$ within $V_{n + 1}^{\mathfrak k, d}$.
Such introduction is very useful in applications,
when one has to deal with operators acting in $L^2 \left( \mathbb R^d \right)$
and admitting the separation of variables
\begin{equation}
    \mathcal T
    =
    \sum_{m = 1}^M
    \alpha_m
    \mathcal T_1
    \otimes \ldots \otimes
    \mathcal T_d
\end{equation}
precisely or approximately.
Here each operator $\mathcal T_j$ acts in $L^2(\mathbb R)$.

Therefore, until the end of this section we stick to the case of
a one dimensional operator $\mathcal T$ and introduce the following matrix elements
\begin{equation}
\label{nonstandard_form_matrices}
\begin{aligned}
    \left[ \sigma_{l'l}^n \right]_{j'j}
    =
    \int_0^1
    \phi_{j'l'}^n(x) \mathcal T \phi_{jl}^n(x)
    dx
    , \qquad
    &
    \left[ \gamma_{l'l}^n \right]_{j'j}
    =
    \int_0^1
    \phi_{j'l'}^n(x) \mathcal T \psi_{jl}^n(x)
    dx
    ,
    \\
    \left[ \beta_{l'l}^n \right]_{j'j}
    =
    \int_0^1
    \psi_{j'l'}^n(x) \mathcal T \phi_{jl}^n(x)
    dx
    , \qquad
    &
    \left[ \alpha_{l'l}^n \right]_{j'j}
    =
    \int_0^1
    \psi_{j'l'}^n(x) \mathcal T \psi_{jl}^n(x)
    dx
    .
\end{aligned}
\end{equation}
For a given scale $n \in \mathbb N_0$,
every symbol
\(
    \alpha^n, \beta^n, \gamma^n
\)
and $\sigma^n$
stands for a matrix of $2^n \times 2^n$-size,
where each element is itself a matrix of $\mathfrak k \times \mathfrak k$-size.
Similarly to \eqref{decomposition_step},
different scales are connected by the wavelet transform 
\begin{equation}
\label{nonstandard_form_decomposition_step}
    \begin{pmatrix}
        \sigma_{l'l}^n
        &
        \gamma_{l'l}^n
        \\
        \beta_{l'l}^n
        &
        \alpha_{l'l}^n
    \end{pmatrix}
    =
    U
    \begin{pmatrix}
        \sigma_{2l', 2l}^{n + 1}
        &
        \sigma_{2l', 2l + 1}^{n + 1}
        \\
        \sigma_{2l' + 1, 2l}^{n + 1}
        &
        \sigma_{2l' + 1, 2l + 1}^{n + 1}
    \end{pmatrix}
    U^T
    .
\end{equation}

For a convolution of the form
\begin{equation}
\label{convolution}
    \mathcal T u (x)
    =
    \int_{\mathbb R}
    T(x - y) u(y)
    dy
\end{equation}
the matrix elements \eqref{nonstandard_form_matrices}
depend on the distance $l' - l$ to the main diagonal:
precisely,
\(
    \left[ \sigma_{l'l}^n \right]_{j'j}
    =
    \left[ \sigma_{l' - l}^n \right]_{j'j}
\)
and so on.
These elements are computed by
\begin{equation}
\label{cross_correlation_convolution_operator_representation}
    \left[ \sigma_l^n \right]_{j'j}
    =
    2^{- n / 2}
    \sum_{p = 0}^{2 \mathfrak k - 1}
    \left(
        c_{j'jp}^{(+)} s_{p, l}^n(T)
        +
        c_{j'jp}^{(-)} s_{p, l - 1}^n(T)
    \right)
    ,
\end{equation}
where the scaling coefficients of the kernel $T$ are defined by \eqref{scaling_coefficients}.
The cross correlation coefficients $c_{j'jp}^{(\pm)}$ are tabulated.
Equation \eqref{cross_correlation_convolution_operator_representation}
gives the pure scaling component of the operator, the first integral in \eqref{nonstandard_form_matrices}.
The other components can be calculated by the decomposition transformation
\eqref{nonstandard_form_decomposition_step} that
simplifies to
\begin{equation}
\label{nonstandard_form_convolution_decomposition_step}
    \begin{pmatrix}
        \sigma_l^n
        &
        \gamma_l^n
        \\
        \beta_l^n
        &
        \alpha_l^n
    \end{pmatrix}
    =
    U
    \begin{pmatrix}
        \sigma_{2l}^{n + 1}
        &
        \sigma_{2l-1}^{n + 1}
        \\
        \sigma_{2l+1}^{n + 1}
        &
        \sigma_{2l}^{n + 1}
    \end{pmatrix}
    U^T
\end{equation}
for the convolution operator \eqref{convolution}.

An important example of the convolution operator \eqref{convolution}
is the exponential heat operator
\(
    \mathcal T = \exp \left( t \partial_x^2 \right)
    ,
\)
for a given time parameter $t > 0$,
having the Gaussian kernel
\(
    T(x) = \alpha_t e^{- \beta_t x^2}
\)
with
$\beta_t = (4t)^{-1}$
and
$\alpha_t = \sqrt{\beta_t / \pi}$.
The scaling coefficients
\(
    s_{p, l}^n(T)
\)
can be calculated easily with high precision.
Then from \eqref{cross_correlation_convolution_operator_representation},
\eqref{nonstandard_form_convolution_decomposition_step}
one obtains the corresponding \ac{NS}-form matrices
\(
    \alpha^n, \beta^n, \gamma^n
    .
\)
Heat kernels possess a narrow, diagonally banded structure in their \ac{NS}-form for small $t > 0$.
Thus, the multiwavelet representation of the heat operator can be considered sparse.

In \cite{Alpert_Beylkin_Gines_Vozovoi2002} a multiresolution of derivative $\partial_x$
was introduced in a weak sense.
In fact, it forms a two-parameter family of derivative representations.
Among them we pick the following
\begin{equation}
\label{weak_derivative_representation}
\begin{aligned}
    \left[ \sigma_1^n \right]_{j'j}
    &
    =
    - 2^{n - 1} \phi_{j'}(0) \phi_j(1)
    ,
    \\
    \left[ \sigma_0^n \right]_{j'j}
    &
    =
    2^{n - 1} \phi_{j'}(1) \phi_j(1) - 2^{n - 1} \phi_{j'}(0) \phi_j(0) - 2^n K_{j'j}
    ,
    \\
    \left[ \sigma_{-1}^n \right]_{j'j}
    &
    =
    2^{n - 1} \phi_{j'}(1) \phi_j(0)
    ,
\end{aligned}
\end{equation}
and
\(
    \left[ \sigma_l^n \right]_{j'j} = 0
\)
for $|l| > 1$.
Note that the derivative can be viewed as a generalized convolution.
Therefore, the multiresolution matrix elements depend on the distance to the diagonal
$l = - 2^n + 1, \ldots, 2^n - 1$,
and so
the corresponding weak derivative \ac{NS}-form matrices
\(
    \alpha^n, \beta^n, \gamma^n
\)
can be found from the wavelet transform \eqref{nonstandard_form_convolution_decomposition_step}.
This representation can be used to form a numerical
representation of Laplacian, that we denote $\Delta_{\text{weak}}$.

In \cite{Anderson_Harrison_Sekino_Sundahl_Beylkin_Fann_Jensen_Sagert2019}
an alternative approximation of derivatives $\partial_x^p$ was introduced.
An auxiliary smooth function $f$ can be projected to piece-wise polynomials
$V_n^{\mathfrak k}$ or interpolated with the same order of accuracy by B-splines
admitting $p$-th order derivatives.
Thus, transformation between the scaling basis $\phi_{jl}^n$ and B-splines,
then between $p$-derivatives of B-splines and back to the scaling basis $\phi_{jl}^n$,
defines an alternative matrix
\(
    \left[ \sigma_{l'l}^n \right]_{j'j}
\)
for the derivative $\partial_x^p$.
The corresponding Laplacian is denoted by $\Delta_{\text{bs}}$.

A computational software based on the described theory is available at \cite{MRChemSoft}.
All numerical experiments presented below are conducted using this software.

\section{Laplacian decomposition}
\label{Laplacian_decomposition_section}
\setcounter{equation}{0}

For given $\tau > 0$ and nonnegative integer $M$, the heat semigroup operator
can be expanded in the Taylor polynomial
\begin{equation}
\label{heat_taylor}
    e^{\tau \Delta}
    =
    \sum_{m = 0}^M
    \frac 1{m!} ( \tau \Delta )^m
    +
    O \left( \tau^{M + 1} \right)
    ,
\end{equation}
which is to be understood in the strong sense.
More precisely, for any sufficiently smooth function $f$, we have the estimate
\begin{equation*}
    \norm
    {
        e^{\tau \Delta} f
        -
        \sum_{m = 0}^M
        \frac 1{m!} ( \tau \Delta )^m f
    }
    _{L^2}
    \leqslant
    \frac {\tau^{M + 1}}{(M + 1)!}
    \norm
    {
        \Delta^{M + 1} f
    }
    _{L^2}
    .
\end{equation*}
In this sense one can approximate the Laplace operator as
\begin{equation}
\label{tau_delta_order_2}
    \tau \Delta
    =
    e^{\tau \Delta} - 1
    +
    O \left( \tau^2 \right)
    .
\end{equation}
Introducing the following notation
\begin{equation}
\label{laplace_order_1}
    \Delta_1(\tau)
    =
    \frac 1{\tau}
    \left(
        e^{\tau \Delta} - 1
    \right)
\end{equation}
one can notice that this operator is bounded in $L^2 \left( \mathbb R^d \right)$
as
\begin{equation}
    \norm
    {
        \tau \Delta_1(\tau)
    }
    =
    \sup_{x \in [0, 1]}
    \left|
        x - 1
    \right|
    =
    1
    ,
\end{equation}
where we have used the fact that the symbol $e^{- \tau |\xi|^2}$
associated with the pseudodifferential operator $e^{\tau \Delta}$
takes all the values in the interval $(0, 1]$.

\begin{figure}[!htbp]
    \centering
    {
        \includegraphics[width=0.7\textwidth]
        {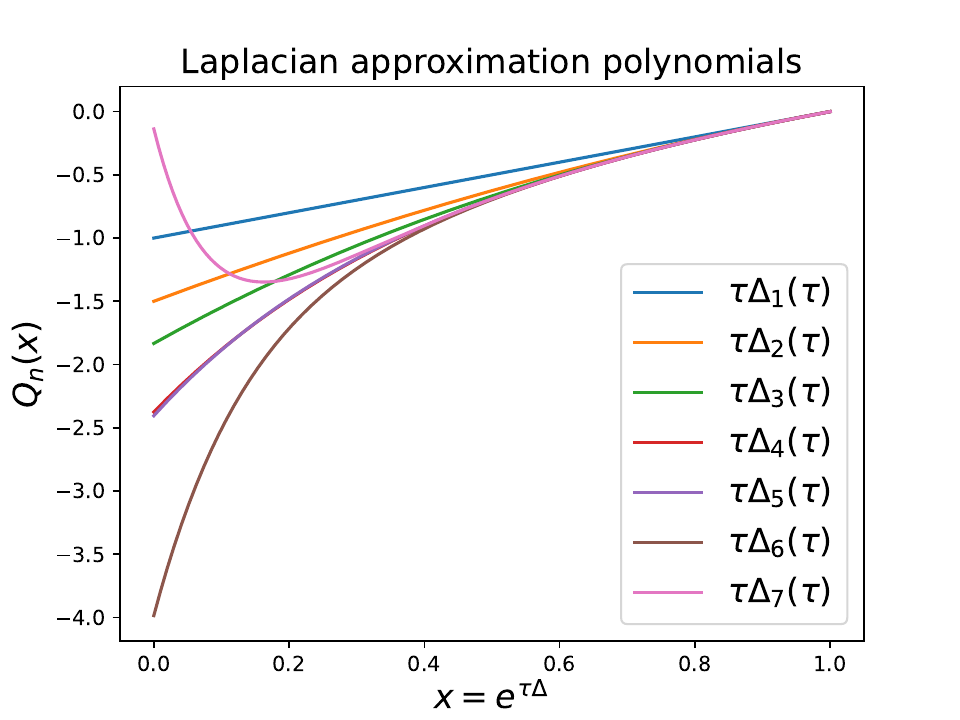}
    }
    \caption
    {
        Polynomials $Q_n(x)$ associated to approximations $\tau \Delta_n(\tau)$.
    }
\label{laplace_polynomial_figure}
\end{figure}

In order to get a more precise expansion than \eqref{tau_delta_order_2},
we consider \eqref{heat_taylor} with two different time parameters, $\tau$ and $2 \tau$,
and subtract them in such a way that $\tau^2$ term is eliminated
\begin{equation*}
    4 e^{\tau \Delta} - e^{2 \tau \Delta}
    =
    3
    +
    2 \tau \Delta
    +
    O \left( \tau^3 \right)
\end{equation*}
and therefore, we have
\begin{equation}
\label{tau_delta_order_3}
    \tau \Delta
    =
    2 e^{\tau \Delta} - \frac 12 e^{2 \tau \Delta} - \frac 32
    +
    O \left( \tau^3 \right)
    .
\end{equation}
In other words,
Laplacian can be approximated by the following expansion
\begin{equation}
\label{laplace_order_2}
    \Delta_2(\tau)
    =
    \frac 1{2\tau}
    \left(
        4 e^{\tau \Delta} - e^{2 \tau \Delta} - 3
    \right)
\end{equation}
bounded as
\begin{equation}
    \norm
    {
        \tau \Delta_2(\tau)
    }
    =
    \sup_{x \in [0, 1]}
    \left|
        2 x - \frac 12 x^2 - \frac 32
    \right|
    =
    \frac 32
    .
\end{equation}

Now having at hand \eqref{laplace_order_1} and \eqref{laplace_order_2},
we can obtain next order approximations
\begin{equation*}
    \Delta
    =
    \Delta_n(\tau)
    +
    O \left( \tau^n \right)
\end{equation*}
recursively as
\begin{equation}
\label{laplace_recursion}
    \tau \Delta_N(\tau)
    =
    \tau \Delta_2(\tau)
    +
    \sum_{m = 3}^N
    \frac{2^{m - 1} - 2}{m!}
    \left(
        \tau \Delta_{N - m + 1}(\tau)
    \right) ^m
    .
\end{equation}
For example,
the third order bounded Laplace operator decomposition is defined by
\begin{equation*}
    \tau \Delta_3(\tau)
    =
    \frac 13 e^{3 \tau \Delta} - \frac 32 e^{2 \tau \Delta} + 3 e^{\tau \Delta} - \frac {11}6
\end{equation*}
with
\begin{equation*}
    \norm
    {
        \tau \Delta_3(\tau)
    }
    =
    \sup_{x \in [0, 1]}
    \left|
        \frac 13 x^3 - \frac 32 x^2 + 3 x - \frac {11}6
    \right|
    =
    \frac {11}6
    .
\end{equation*}
Clearly, every operator $\tau \Delta_n(\tau)$ is a polynomial $Q_n(x)$
in the variable $x = e^{\tau \Delta}$,
with coefficients independent of $\tau$.
Note that the Laplacian recursive algorithm \eqref{laplace_recursion} guarantees that we have minimal possible
degree $\deg Q_n$, while increasing the approximation order $n \in \mathbb N$.
Nevertheless, the degree increases rapidly with $n$,
and so it does not seem practical to consider $n > 9$.
In Section \ref{Free_particle_propagator_section} we will need to know
the norms for these approximations.
Therefore, we summarize this information in Table \ref{laplace_degree_norm_table}.
The norms for $n = 7, 8, 9$ are given approximately.
The first seven polynomials are also plotted in Figure \ref{laplace_polynomial_figure}.

\begin{table}[h!]
    \centering
    \renewcommand{\arraystretch}{1.4} 
    \begin{tabular}{|c|c|c|c|c|c|c|c|c|c|}
        \hline
         $n$        & 1 & 2 & 3 & 4 & 5 & 6  & 7  & 8  & 9  \\ 
        \hline
         $\deg Q_n$ & 1 & 2 & 3 & 6 & 9 & 18 & 27 & 54 & 81 \\ 
        \hline
         $\norm{\tau \Delta_n(\tau)}$ &
         1 & $\frac 32$ & $\frac {11}6$ & $\frac {19}8$ &
         $\frac {62339}{25920}$ & $\frac {826433}{207360}$ &
         1.348 & 21.68 & 58.89 \\ 
        \hline
    \end{tabular}
    \caption{
        Polynomial approximation of Laplacian
        $\tau \Delta \approx \tau \Delta_n(\tau) = Q_n \left( e^{\tau \Delta} \right)$.
    }
    \label{laplace_degree_norm_table}
\end{table}

Finally, it remains to assess the performance of the introduced Laplace operator decompositions.
We test it on the function
\begin{equation}
\label{derivative_test_function_f}
    f(x)
    =
    C
    \exp \left( - \frac {|x - x_0|^2} {\sigma} \right)
    , \quad
    x \in \mathbb R^d
    ,
\end{equation}
transforming to
\begin{equation}
\label{derivative_test_function_g}
    g(x)
    =
    \Delta f(x)
    =
    \frac C{\sigma}
    \left( \frac 4{\sigma} |x - x_0|^2 - 2d \right)
    \exp \left( - \frac {|x - x_0|^2} {\sigma} \right)
    .
\end{equation}
There is an obvious relation
\begin{equation*}
    \norm g_{L^2}
    =
    \frac C2
    \pi^{d/4}
    \left( \frac {\sigma}2 \right)^{d/4 - 1}
    \sqrt{
        d (d + 2)
    }
\end{equation*}
between an arbitrary constant $C$ and the norm of second derivative $g$.
One can choose $C$ so that $\norm g_{L^2} = 1$, for example.
Operators $e^{m \tau \Delta}$ with $m \in \mathbb N$ are regarded as convolutions
\eqref{convolution} and discretized by 
\eqref{cross_correlation_convolution_operator_representation},
\eqref{nonstandard_form_convolution_decomposition_step}.





\begin{table}[h!]
\centering
\renewcommand{\arraystretch}{1.4}
\begin{tabular}{|c|c|c|c|c|c|c|c|}
\hline
 & & \multicolumn{6}{c|}{$\Delta$} \\
 \cline{3-8}
 & & $\Delta_{4}$ & $\Delta_{5}$ & $\Delta_{6}$ & $\Delta_{7}$ & $\Delta_{\text{{weak}}}$ & $\Delta_{\text{{bs}}}$ \\
\hline
\multirow{2}{*}{$C = 1$} & $\mathfrak k = 8$ & $1.7 \cdot 10^{-7}$ & $1.7 \cdot 10^{-7}$ & $2.1 \cdot 10^{-7}$ & $7.0 \cdot 10^{-9}$ & $2.1 \cdot 10^{-8}$ & $1.9 \cdot 10^{-9}$ \\
 & $\mathfrak k = 15$ & $7.3 \cdot 10^{-8}$ & $5.3 \cdot 10^{-8}$ & $7.8 \cdot 10^{-8}$ & $5.8 \cdot 10^{-8}$ & $2.7 \cdot 10^{-8}$ & $1.3 \cdot 10^{-5}$ \\
\hline
\multirow{2}{*}{$\norm g_{L^2} = 1$} & $\mathfrak k = 8$ & $8.2 \cdot 10^{-7}$ & $8.2 \cdot 10^{-7}$ & $1.5 \cdot 10^{-6}$ & $4.2 \cdot 10^{-7}$ & $7.0 \cdot 10^{-6}$ & $4.1 \cdot 10^{-7}$ \\
 & $\mathfrak k = 15$ & $1.1 \cdot 10^{-7}$ & $7.2 \cdot 10^{-8}$ & $1.4 \cdot 10^{-7}$ & $6.1 \cdot 10^{-8}$ & $1.2 \cdot 10^{-7}$ & $1.3 \cdot 10^{-5}$ \\
\hline
\end{tabular}
\caption{
Error $\norm{ g - \Delta_{\text{numer}} f }_{L^2} / \norm{ g }_{L^2}$
with \eqref{derivative_test_function_f}, \eqref{derivative_test_function_g}
in the dimension $d = 1$
for $\sigma = 10^{-5}, \tau = 10^{-8}$ 
and precision $\varepsilon = 10^{-9}$.
}
\label{laplace_accuracy_1d_table}
\end{table}





\begin{table}[h!]
\centering
\renewcommand{\arraystretch}{1.4}
\begin{tabular}{|c|c|c|c|c|c|c|c|}
\hline
 & & \multicolumn{6}{c|}{$\Delta$} \\
 \cline{3-8}
 & & $\Delta_{1}$ & $\Delta_{2}$ & $\Delta_{3}$ & $\Delta_{4}$ & $\Delta_{\text{{weak}}}$ & $\Delta_{\text{{bs}}}$ \\
\hline
\multirow{2}{*}{$C = 1$} & $\mathfrak k = 8$ & $4.0 \cdot 10^{-3}$ & $3.1 \cdot 10^{-5}$ & $3.7 \cdot 10^{-7}$ & $3.1 \cdot 10^{-8}$ & $8.1 \cdot 10^{-7}$ & $4.6 \cdot 10^{-8}$ \\
 & $\mathfrak k = 15$ & $4.0 \cdot 10^{-3}$ & $3.1 \cdot 10^{-5}$ & $3.7 \cdot 10^{-7}$ & $1.3 \cdot 10^{-8}$ & $3.8 \cdot 10^{-8}$ & $6.3 \cdot 10^{-6}$ \\
\hline
\multirow{2}{*}{$\norm g_{L^2} = 1$} & $\mathfrak k = 8$ & $4.0 \cdot 10^{-3}$ & $3.1 \cdot 10^{-5}$ & $1.0 \cdot 10^{-6}$ & $1.4 \cdot 10^{-6}$ & $6.4 \cdot 10^{-6}$ & $3.8 \cdot 10^{-7}$ \\
 & $\mathfrak k = 15$ & $4.0 \cdot 10^{-3}$ & $3.1 \cdot 10^{-5}$ & $3.7 \cdot 10^{-7}$ & $7.3 \cdot 10^{-8}$ & $8.2 \cdot 10^{-7}$ & $6.3 \cdot 10^{-6}$ \\
\hline
\end{tabular}
\caption{
Error $\norm{ g - \Delta_{\text{numer}} f }_{L^2} / \norm{ g }_{L^2}$
with \eqref{derivative_test_function_f}, \eqref{derivative_test_function_g}
in the dimension $d = 3$
for $\sigma = 10^{-4}, \tau = 10^{-7}$ 
and precision $\varepsilon = 10^{-9}$.
}
\label{laplace_accuracy_3d_table}
\end{table}


When working with multiwavelets,
a truncation precision parameter $\varepsilon > 0$ must be introduced.
This parameter quantifies how well the multiresolution of functions and operators,
adaptively truncated at specific scaling levels $n$,
approximates the corresponding exact entities.
In essence, $\varepsilon$ serves as a reliable upper bound for the approximation error in the $L^2$-norm
for most chemistry applications \cite{Frediani_Fossgaard_Fla_Ruud}.
In the tests below for illustrative purposes,
we use a high precision value of $\varepsilon = 10^{-9}$
to ensure that the truncation error does not interfere with the error from the heat semigroup approximation.

\begin{figure}[ht!]
\centering
\begin{subfigure}{0.49\textwidth}
\label{subfig_Heat_polynom7_order4}
	\includegraphics[width=\textwidth]{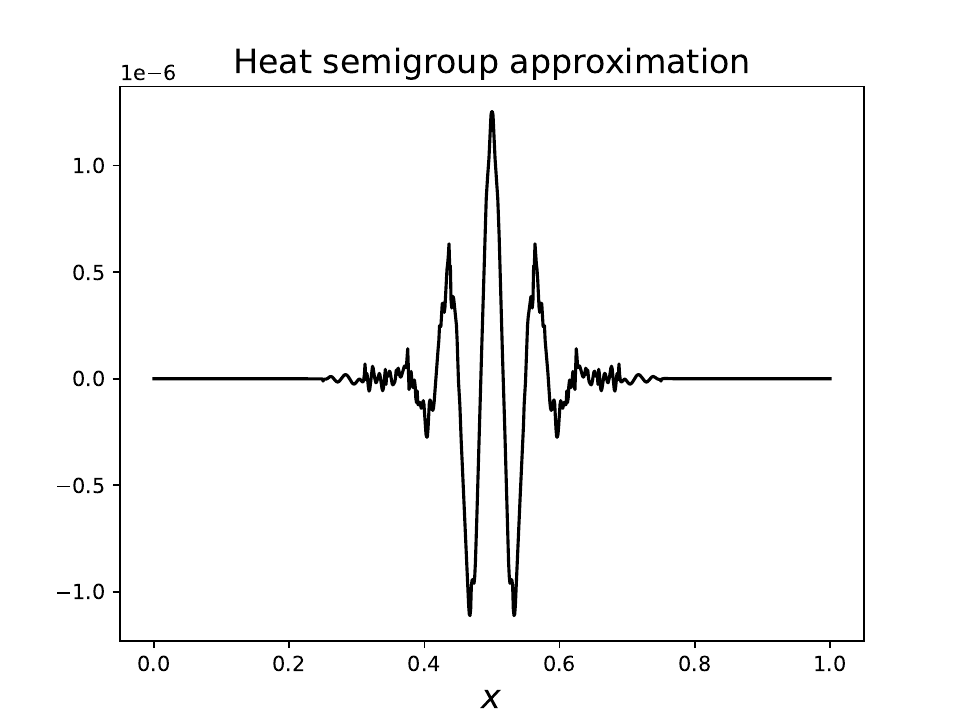}
	\caption{$\Delta_{\text{numer}} = \Delta_4(\tau)$, $\text{error} \approx 2.6 \cdot 10^{-7}$}
\end{subfigure}
\hfill
\begin{subfigure}{0.49\textwidth}
\label{subfig_Heat_polynom7_order7}
	\includegraphics[width=\textwidth]{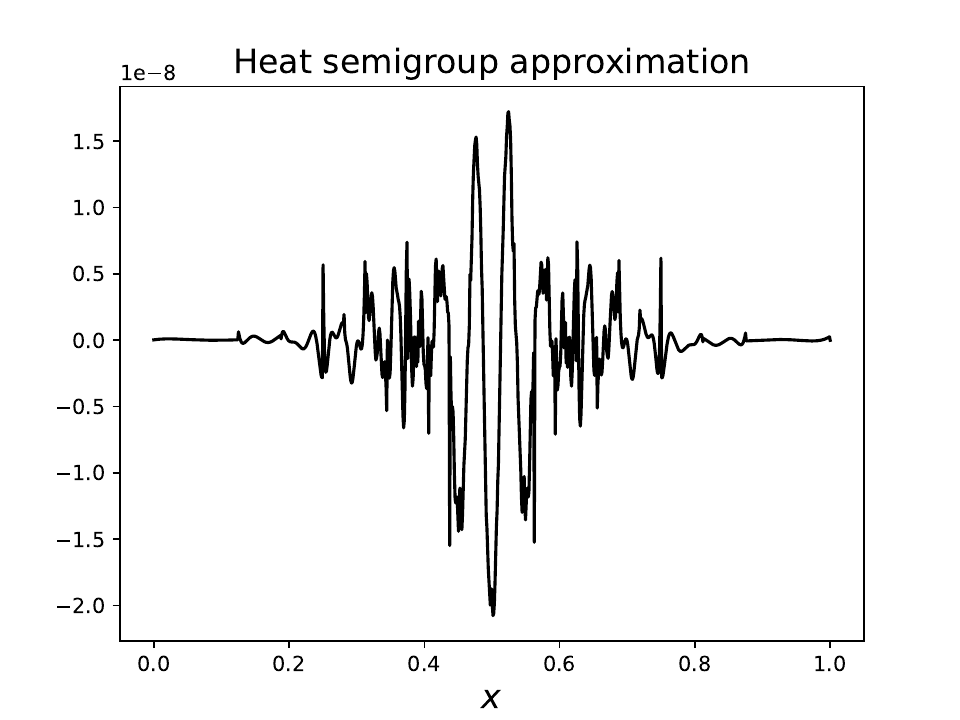}
	\caption{$\Delta_{\text{numer}} = \Delta_7(\tau)$, $\text{error} \approx 4.4 \cdot 10^{-9}$}
\end{subfigure}
\hfill
\begin{subfigure}{0.49\textwidth}
\label{subfig_BS_polynom7}
	\includegraphics[width=\textwidth]{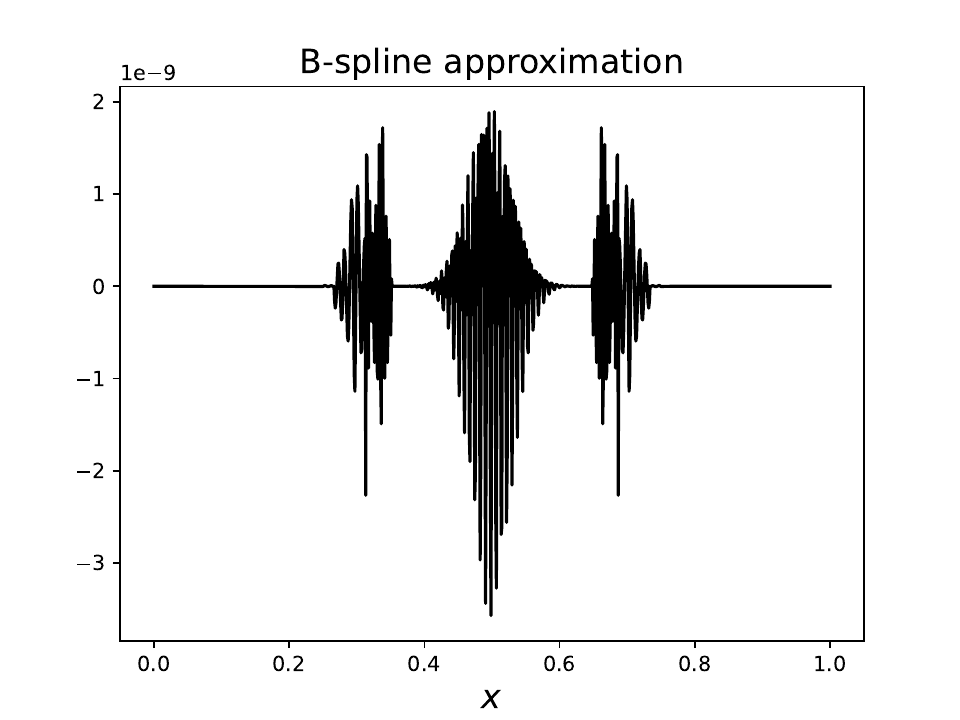}
	\caption{$\Delta_{\text{numer}} = \Delta_{\text{bs}}$, $\text{error} \approx 5.2 \cdot 10^{-10}$}
\end{subfigure}
\hfill
\begin{subfigure}{0.49\textwidth}
\label{subfig_ABGV_polynom7}
	\includegraphics[width=\textwidth]{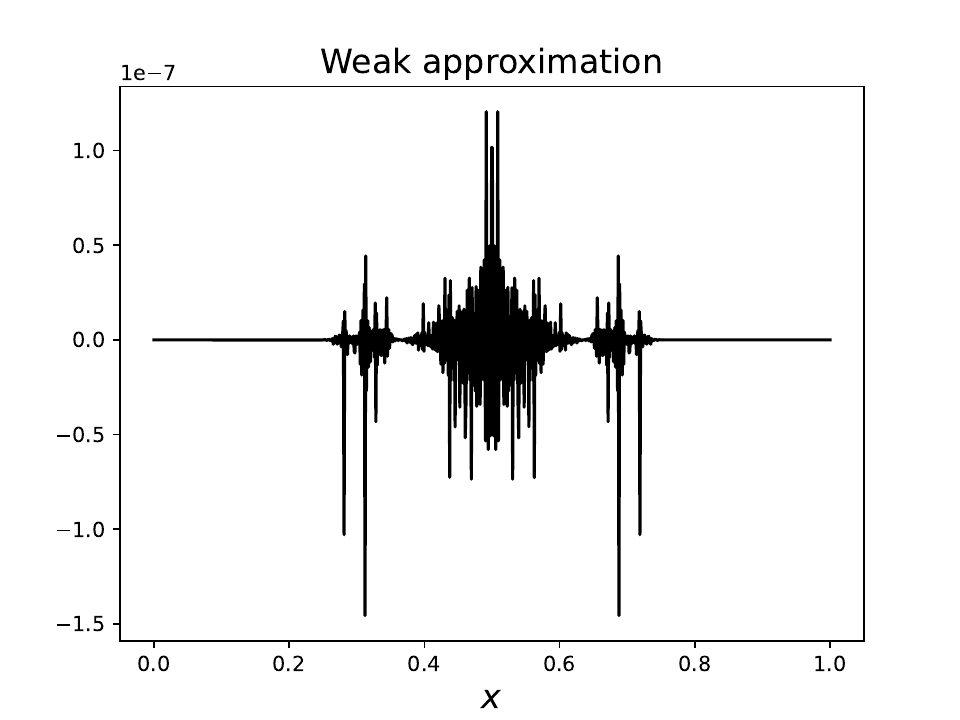}
	\caption{$\Delta_{\text{numer}} = \Delta_{\text{weak}}$, $\text{error} \approx 1.8 \cdot 10^{-8}$}
\end{subfigure}
\hfill
\caption
{
    The relative difference $( g(x) - \Delta_{\text{numer}} f(x) )/ \norm{g}$
    for the \ac{MRA} order $\mathfrak k = 8$ in the one dimensional case $d = 1$.
    Here $C = 1$, $\sigma = 0.002$ in \eqref{derivative_test_function_f},
    approximation parameter $\tau = 10^{-5}$
    and precision $\varepsilon = 10^{-9}$.
}
\label{figure_laplace_error_polynom7}
\end{figure}
\begin{figure}[ht!]
\centering
\begin{subfigure}{0.49\textwidth}
\label{subfig_Heat_polynom14_order4}
	\includegraphics[width=\textwidth]{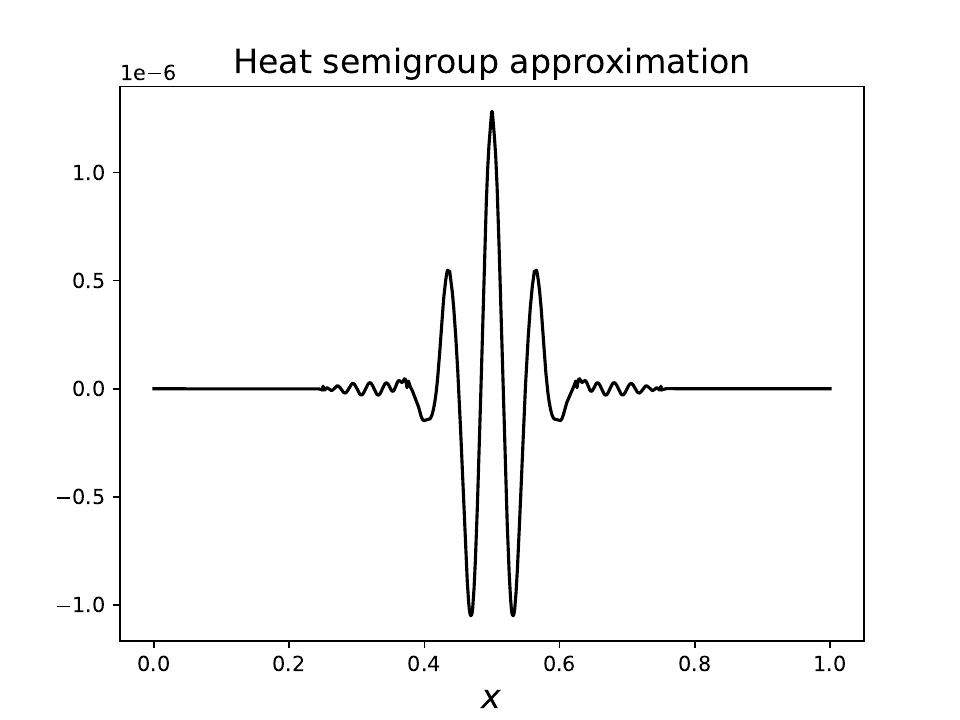}
	\caption{$\Delta_{\text{numer}} = \Delta_4(\tau)$, $\text{error} \approx 2.6 \cdot 10^{-7}$}
\end{subfigure}
\hfill
\begin{subfigure}{0.49\textwidth}
\label{subfig_Heat_polynom14_order7}
	\includegraphics[width=\textwidth]{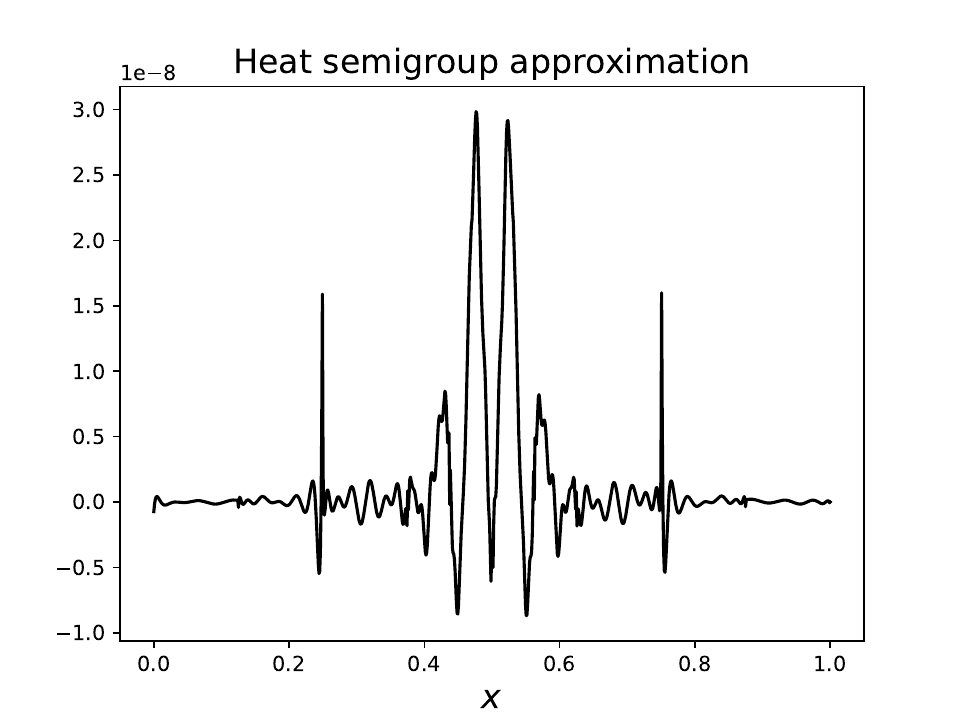}
	\caption{$\Delta_{\text{numer}} = \Delta_7(\tau)$, $\text{error} \approx 5.4 \cdot 10^{-9}$}
\end{subfigure}
\hfill
\begin{subfigure}{0.49\textwidth}
\label{subfig_BS_polynom14}
	\includegraphics[width=\textwidth]{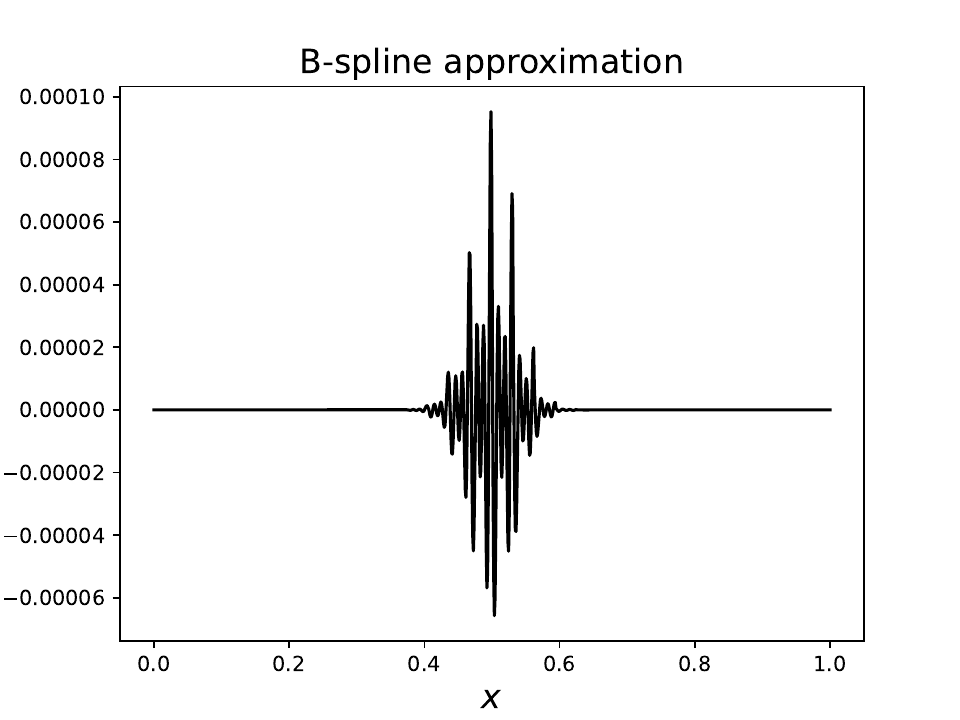}
	\caption{$\Delta_{\text{numer}} = \Delta_{\text{bs}}$, $\text{error} \approx 1.0 \cdot 10^{-5}$}
\end{subfigure}
\hfill
\begin{subfigure}{0.49\textwidth}
\label{subfig_ABGV_polynom14}
	\includegraphics[width=\textwidth]{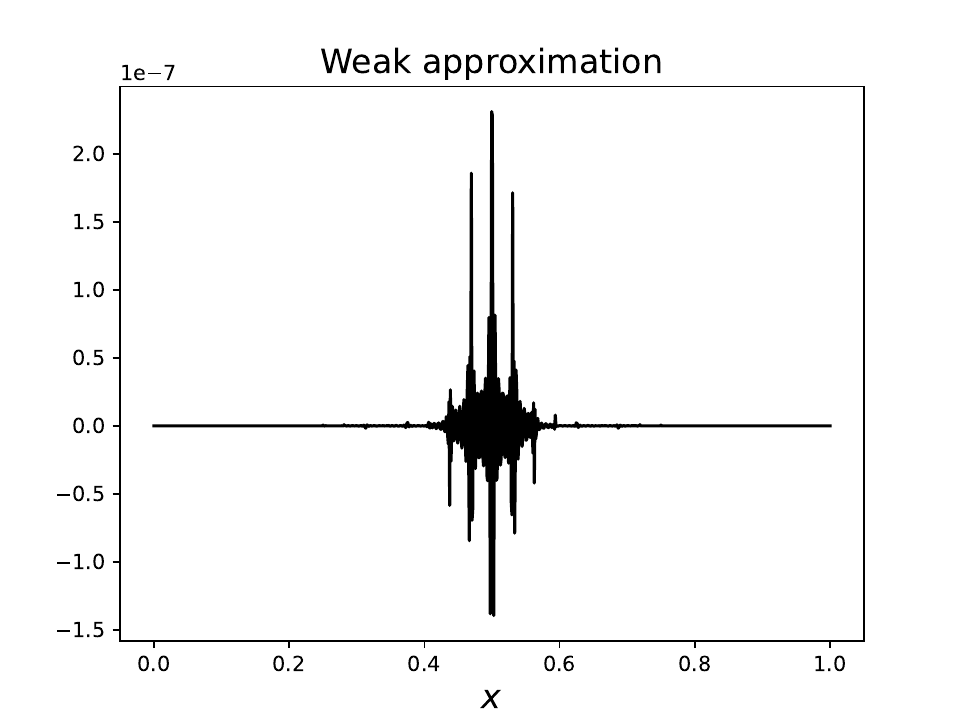}
	\caption{$\Delta_{\text{numer}} = \Delta_{\text{weak}}$, $\text{error} \approx 2.2 \cdot 10^{-8}$}
\end{subfigure}
\hfill
\caption
{
    The relative difference $( g(x) - \Delta_{\text{numer}} f(x) )/ \norm{g}$
    for the \ac{MRA} order $\mathfrak k = 15$ in the one dimensional case $d = 1$.
    Here $C = 1$, $\sigma = 0.002$ in \eqref{derivative_test_function_f},
    approximation parameter $\tau = 10^{-5}$
    and precision $\varepsilon = 10^{-9}$.
}
\label{figure_laplace_error_polynom14}
\end{figure}

\begin{figure}[ht!]
\centering
\begin{subfigure}{0.49\textwidth}
    \includegraphics[width=\textwidth]{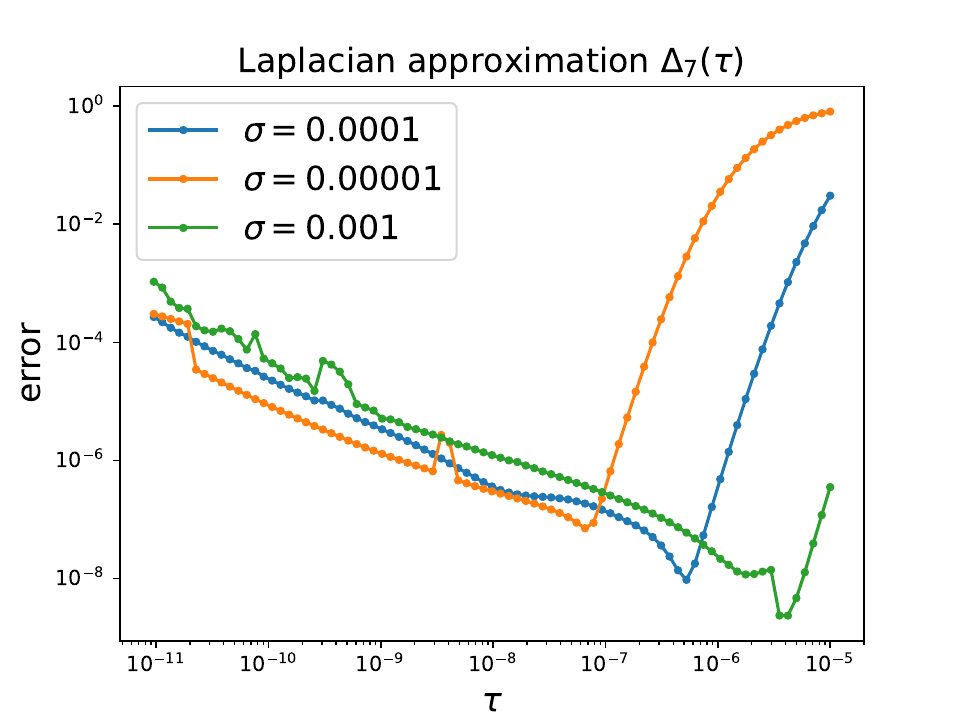}
    \caption{
        $d = 1$, $n = 7$
    }
\end{subfigure}
\hfill
\begin{subfigure}{0.49\textwidth}
    \includegraphics[width=\textwidth]{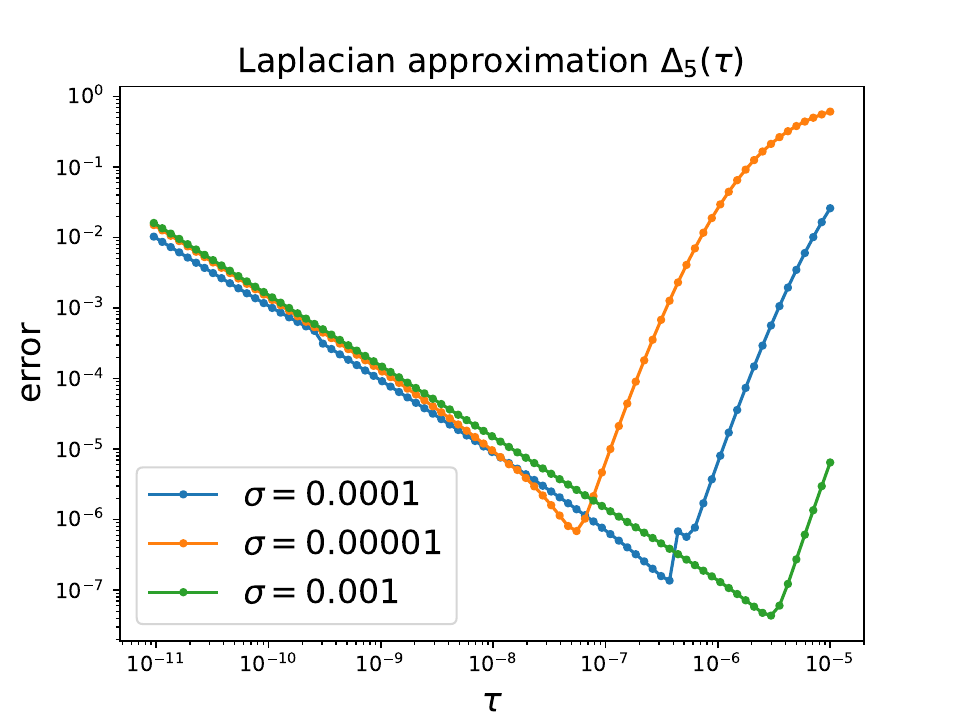}
    \caption{
        $d = 3$, $n = 5$
    }
\end{subfigure}
\hfill
\caption
{
    Accuracy in $L^2$-norm of numerical evaluation $g \approx \Delta_n(\tau) f$
    with the test functions $f, g$ defined by \eqref{derivative_test_function_f}, \eqref{derivative_test_function_g}.
    Here $\mathfrak k = 10$, $\norm g_{L^2} = 1$
    and precision $\varepsilon = 10^{-9}$.
}
\label{laplace_error_dependence_on_tau_figure}
\end{figure}

The accuracy of performance of different Laplacian approximations
for several different test parameters
are summarized in Table \ref{laplace_accuracy_1d_table} for $d = 1$
and Table \ref{laplace_accuracy_3d_table} for $d = 3$.
Qualitatively, the results are independent of the space dimension $d$.
The introduced here second derivative calculations
are either inferior or superior to $\Delta_{\text{bs}}$ or $\Delta_{\text{weak}}$
introduced at the end of Section \ref{Multiresolution_analysis_section}.
The conclusion depends significantly on the polynomial \ac{MRA} order $\mathfrak k$,
the time parameter $\tau$ and the order $n$ of Laplacian approximation $\Delta \approx \Delta_n(\tau)$.
The most optimal time parameter is $\tau = 10^{-5}$ for $\sigma = 0.002$, for instance.
The corresponding error plots are provided in Figure \ref{figure_laplace_error_polynom7}
for $\mathfrak k = 8$ and in Figure \ref{figure_laplace_error_polynom14} for $\mathfrak k = 15$.
In this particular example our Laplacian $\Delta_n(\tau)$ is not significantly affected by the \ac{MRA} order,
whereas for B-spline and weak derivatives it can be more important.
Indeed, $\Delta_{\text{bs}}$ performs better for small polynomial order $\mathfrak k$
and worse for large $\mathfrak k$, as one can notice in
Tables \ref{laplace_accuracy_1d_table}, \ref{laplace_accuracy_3d_table}
and Figures \ref{figure_laplace_error_polynom7}, \ref{figure_laplace_error_polynom14}.
It is also worth to point out that for larger $\mathfrak k$
the heat approximation demonstrates slightly smoother error in Figure \ref{figure_laplace_error_polynom14}
than for smaller $\mathfrak k$ in Figure \ref{figure_laplace_error_polynom7}.
In particular, this smooth error dependence on the space variable $x \in \mathbb R^d$
suggests that the new operator $\Delta_n(\tau)$ can be implemented iteratively,
in contrast to $\Delta_{\text{bs}}$ and $\Delta_{\text{weak}}$.
This property is the primary motivation for introducing the heat semigroup representation.
The feasibility of iterative implementation of $\Delta_n(\tau)$
is explored in the next two sections in the context of quantum mechanics applications.

\begin{figure}[ht!]
    \centering
    \begin{subfigure}{0.49\textwidth}
        \includegraphics[width=\textwidth]{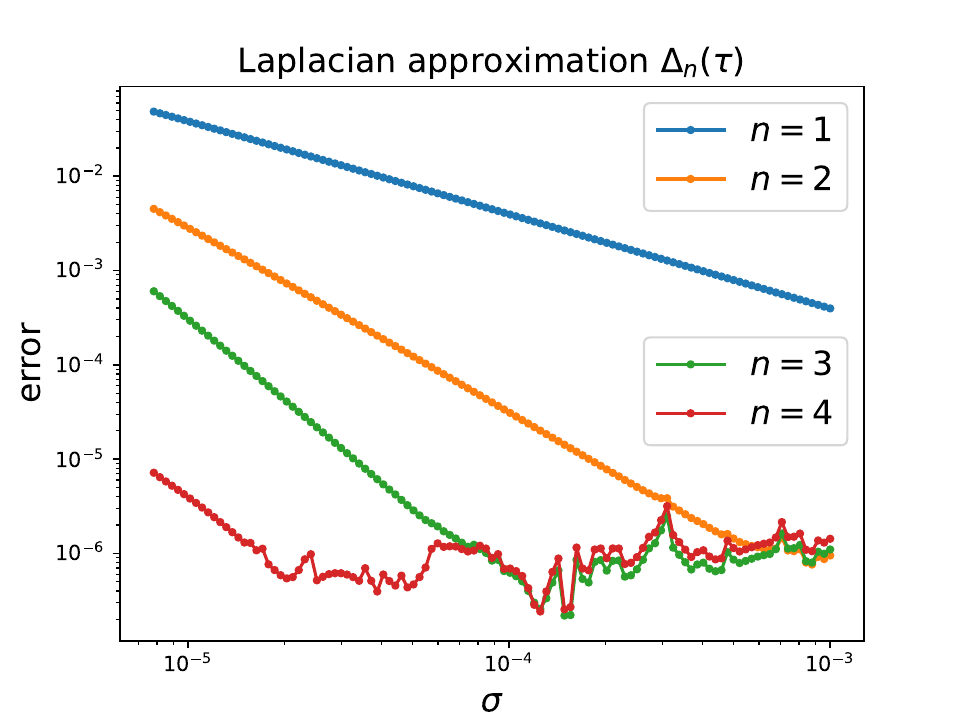}
    \end{subfigure}
    \hfill
    \begin{subfigure}{0.49\textwidth}
        \includegraphics[width=\textwidth]{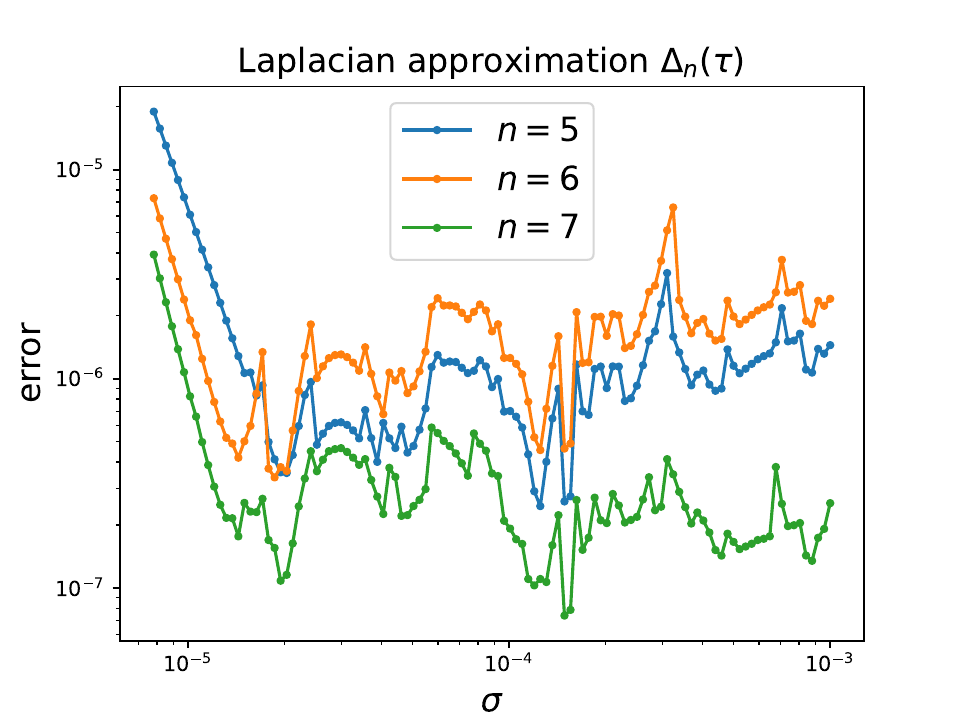}
    \end{subfigure}
    \hfill
    \caption
    {
        The relative difference $\norm{ g - \Delta_{\text{numer}} f} / \norm{g}$
        dependence on the width $\sigma$ of the test function $f$
        defined by \eqref{derivative_test_function_f} in the dimension $d = 3$.
        Here $\mathfrak k = 10$, approximation parameter $\tau = 10^{-7}$,
        $\norm g_{L^2} = 1$ and the threshold $\varepsilon = 10^{-9}$.
    }
\label{laplace_error_dependence_on_sigma_figure}
\end{figure}

A notable drawback of the approximation $\Delta \approx \Delta_n(\tau)$
is the need to tune the temporal parameter $\tau$ to achieve the desired accuracy,
as illustrated in Figure \ref{laplace_error_dependence_on_tau_figure}.
Nevertheless, the author believes that in each practical electronic structure calculation,
it is possible to select $\tau$ based on a characteristic parameter $\sigma$,
which defines the average width of atomic orbitals. 
This hypothesis is further explored in the next section and is partially supported
by Figure \ref{laplace_error_dependence_on_sigma_figure}.
Moreover, in electron dynamics simulations, precise tuning is less critical,
as the error in the Laplacian approximation is suppressed by the small time step $t$.
This is evident from the expression
\(
    e^{i t \Delta}
    \approx
    1 + i t \Delta_n(t)
    .
\)
This aspect is elaborated further in Section \ref{Free_particle_propagator_section}.

\section{Self-consistent field}
\label{Self_consistent_field_section}
\setcounter{equation}{0}

A very simple chemistry ground state problem can be formulated as
\begin{equation}
	\Delta \psi
	=
	2 (V(x) - E) \psi
\end{equation}
that we can formally approximate in the following way
\begin{equation*}
    \Delta_n(\tau) \psi
    =
    2 (V(x) - E) \psi
\end{equation*}
Now recalling
\begin{equation}
\label{recalling_delta}
    \tau \Delta_n(\tau) = a_n + T_n(\tau)
    ,
\end{equation}
where we have separated the identity from the heat propagator at positive times,
one obtains
\begin{equation*}
    \psi
    =
    -
    \frac{ T_n(\tau) }{ a_n } \psi
    +
    \frac{ 2 \tau }{ a_n } (V(x) - E) \psi
    .
\end{equation*}
This eigenvalue problem is
complemented by the normalization condition
\begin{equation*}
    \langle \psi, \psi \rangle = 1
    .
\end{equation*}
From the last two equations we can deduce
\begin{equation*}
    E
    =
    \left \langle \psi \left| - \frac 12 \Delta_n(\tau) + V \right| \psi \right \rangle
    =
    -
    \frac {a_n}{2\tau}
    \left \langle \psi \left| 1 + \frac {T_n(\tau)} {a_n} \right| \psi \right \rangle
    +
    \left \langle \psi \left| V \right| \psi \right \rangle
    .
\end{equation*}
Thus,
we have arrived to an SCF type system
\begin{equation}
\label{SCF_type_system}
    \psi = \mathfrak F(\psi, E)
    , \quad
    E = \mathfrak G(\psi)
\end{equation}
that one can try to solve iteratively
\begin{equation}
\label{simple_iteration}
    \psi_{k + 1} = \mathfrak F(\psi_k, E_k)
    , \quad
    E_k = \mathfrak G(\psi_k)
    .
\end{equation}

The procedure \eqref{simple_iteration} may converge prohibitively slowly.
Therefore, we apply the DIIS acceleration algorithm \cite{Pulay1980},
that for the fixed point problem
\[
    x = \mathfrak f(x)
\]
in a Hilbert space $L$,
is formulated as follows.
Let
\(
    \mathfrak g(x) = x - \mathfrak f(x)
\)
be the residual mapping.
Given a tolerance $\epsilon > 0$ one searches for 
solution of $\mathfrak g(x) = 0$.
One starts from an arbitrary value $x_0 \in L$, $k = 0$
and then iterates over $k$:
\begin{enumerate}
    \item
    Evaluate the residual $r_k = - \mathfrak g(x_k)$.
    \item
    Terminate if the desired precision $\norm{r_k} < \epsilon$ is reached.
    \item 
    Consider a subset of previous iterates
    \(
        x_{\ell(k)}, \ldots, x_k
    \)
    such that the residuals
    \(
        \mathfrak g \left( x_{\ell(k)} \right), \ldots, \mathfrak g(x_k)
    \)
    are linearly independent.
    \item
    Find real numbers
    \(
        c_{\ell(k)}, \ldots, c_k
    \)
    minimizing the norm
    \(
        \norm{\sum_{i = \ell(k)}^k c_i r_i}
    \)
    subject to the constraint
    \(
        \sum_{i = \ell(k)}^k c_i = 1
        .
    \)
    \item
    Introduce the next iterate
    \(
        x_{k + 1} = \sum_{i = \ell(k)}^k c_i \mathfrak f(x_i)
        .
    \)
\end{enumerate}
The convergence of DIIS was analyzed in \cite{Rohwedder_Schneider2011}.
There is some flexibility in defining the mapping $\mathfrak f$
for the SCF problem \eqref{SCF_type_system}:
\begin{enumerate}[label=(SCF.\arabic*), leftmargin=5.5em]
    \item
    \label{SCF_formulation00}
    Let
    $x = (\psi, E) \in L^2 \left( \mathbb R^d \right) \times \mathbb R$
    and
    $\mathfrak f(x) = \left( \mathfrak F(\psi, E),  \mathfrak G(\mathfrak F(\psi, E)) \right)$.
    The procedure is defined by the $L^2$-product norm $\lVert x \rVert = \lVert (\psi, E) \rVert$.
    \item
    \label{SCF_formulation01}
    Let
    $x = (\psi, E) \in L^2 \left( \mathbb R^d \right) \times \mathbb R$
    and
    $\mathfrak f(x) = \left( \mathfrak F(\psi, E),  \mathfrak G(\mathfrak F(\psi, E)) \right)$.
    The procedure is defined by the seminorm $\lVert x \rVert = \lVert \psi \rVert_{L^2}$.
    \item
    \label{SCF_formulation02}
    Let $x = \psi \in L^2 \left( \mathbb R^d \right)$
    and
    $\mathfrak f(x) = \mathfrak F(\psi, \mathfrak G(\psi))$.
\end{enumerate}

\subsection{Smooth potential well}

\begin{figure}[ht!]
    \centering
    \begin{subfigure}{0.49\textwidth}
        \includegraphics[width=\textwidth]{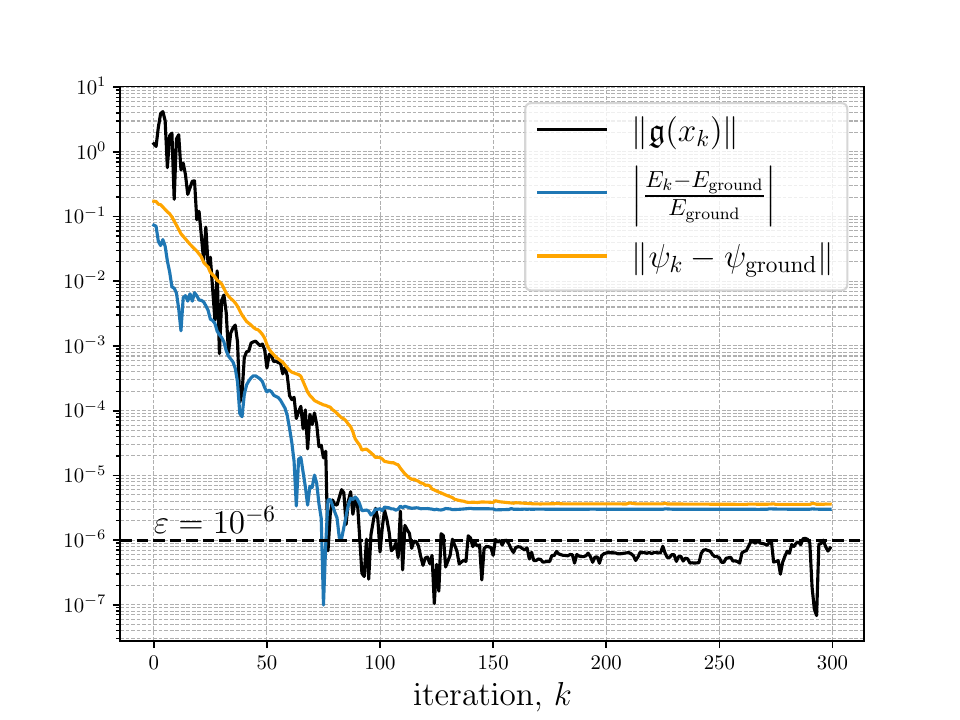}
        \caption{
            Formulation \ref{SCF_formulation00}
        }
    \end{subfigure}
    \hfill
    \begin{subfigure}{0.49\textwidth}
        \includegraphics[width=\textwidth]{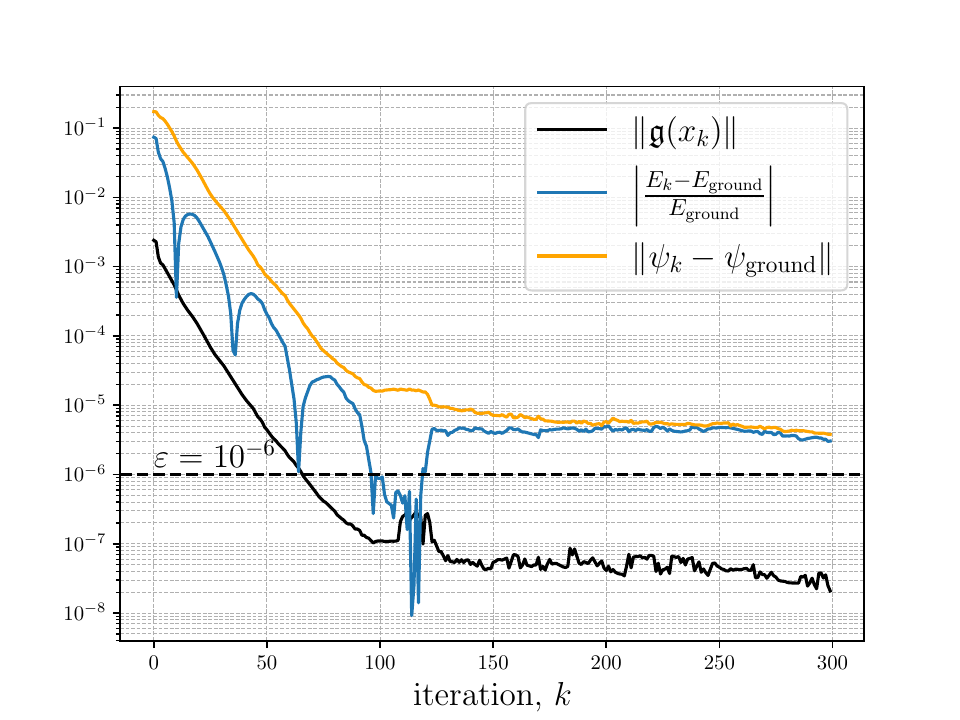}
        \caption{
            Formulation \ref{SCF_formulation02}
        }
    \end{subfigure}
    \hfill
    \caption
    {
        Convergence of DIIS for the smooth potential well in the dimension $d = 1$.
        The second order derivative $\partial_x^2$ is approximated by $\Delta_4(\tau)$
        with $\tau = 10^{-5}$ and the \ac{MRA} order
        $\mathfrak k = 10$ and threshold $\varepsilon = 10^{-6}$.
    }
\label{well_convergence_figure}
\end{figure}

We test the proposed approach on the following one-dimensional Schrödinger equation
\begin{equation*}
    - \frac 12 \partial_x^2 \psi
    + V(x) \psi
    =
    E \psi
    ,
\end{equation*}
defined on the real line $\mathbb R$.
The potential $V(x)$ is given by
$$
    V(x)
    =
    -
    \frac {3 L^2}{ 2 \cosh^2{2L (x - 1/2) } }
$$
The problem is normalised by
$$
    \int_{\mathbb R} |\psi(x)|^2 dx = 1
    .
$$
There is a unique non-trivial solution
$E = E_{\text{ground}} = - L^2/2$
associated to
\[
    \psi(x)
    =
    \psi_{\text{ground}}(x)
    =
    \sqrt{
        \frac {2L}{ \pi \cosh{2L(x - 1/2)} }
    }
\]
that we refer to as the ground state solution.
An auxiliary parameter $L$ is set to $30$,
ensuring that $\psi_{\text{ground}}$ is negligible outside of the numerical domain $[0, 1]$.
As an initial guess $\psi_{\text{guess}}$, we take the function given in \eqref{derivative_test_function_f}
with $\sigma = 0.0064$ and a constant $C > 0$ chosen to ensure that $\psi_{\text{guess}}$ is normalized.
The Laplace operator $\Delta$
is approximated by the fourth order heat semigroup expansion $\Delta_4(\tau)$
with $\tau = 10^{-5}$.
The \ac{MRA} parameters are set as $\mathfrak k = 10$ and $\varepsilon = 10^{-6}$.
We initialize $E_{\text{guess}} = \mathfrak G(\psi_{\text{guess}})$
and perform one iteration of \eqref{simple_iteration} to define the starting values
\begin{equation}
\label{starting_values}
    \psi_0 = \mathfrak F(\psi_{\text{guess}}, E_{\text{guess}})
    , \quad
    E_0 = \mathfrak G(\psi_0)
    .
\end{equation}
Subsequently, we employ the DIIS accelerator using the formulations
\ref{SCF_formulation00}-\ref{SCF_formulation02}
with $0 \leqslant k - \ell(k) \leqslant 11$.
The convergence is demonstrated in Figure \ref{well_convergence_figure}
for formulations \ref{SCF_formulation00} and \ref{SCF_formulation02}.


%
%
%
%

\subsection{Hydrogen atom}

\begin{figure}[ht!]
    \centering
    \begin{subfigure}{0.49\textwidth}
        \includegraphics[width=\textwidth]{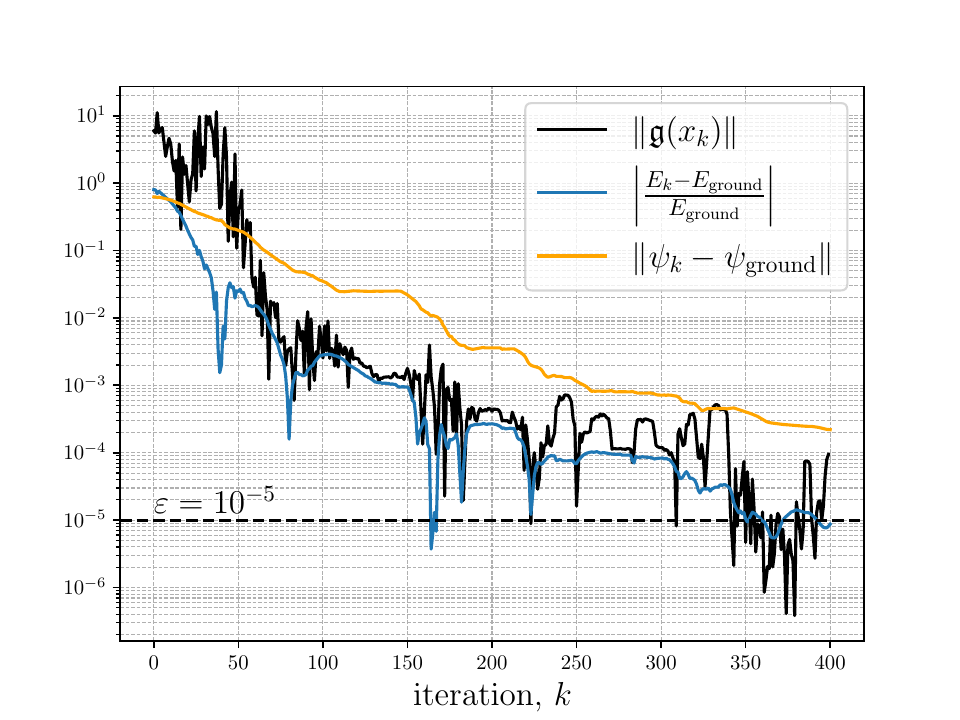}
        \caption{
            Formulation \ref{SCF_formulation00}
        }
    \end{subfigure}
    \hfill
    \begin{subfigure}{0.49\textwidth}
        \includegraphics[width=\textwidth]{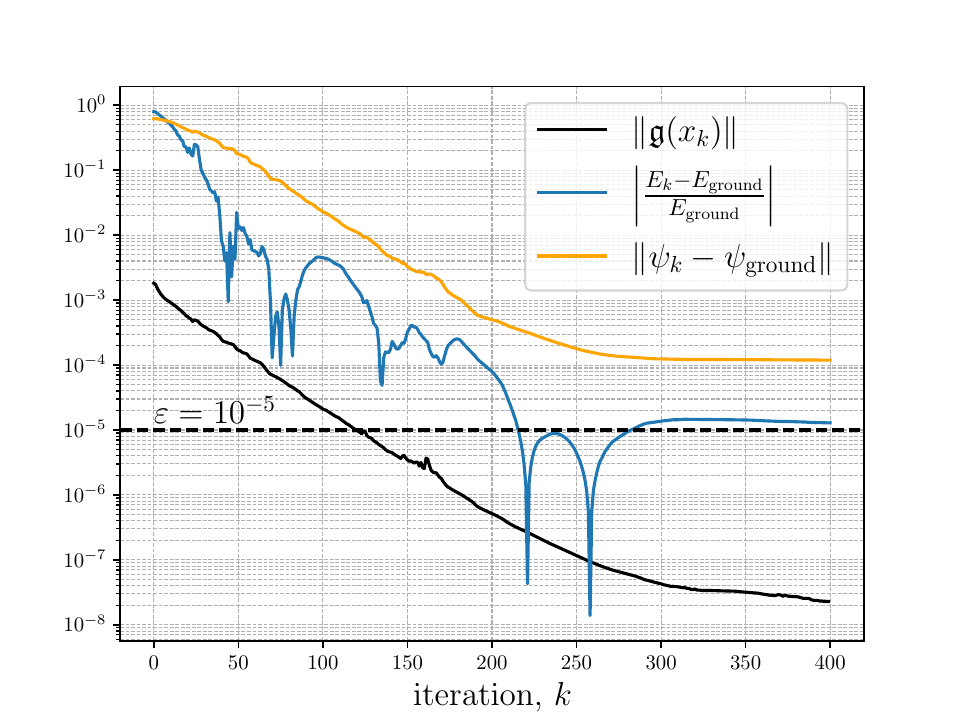}
        \caption{
            Formulation \ref{SCF_formulation01}
        }
    \end{subfigure}
    \hfill
    \begin{subfigure}{0.49\textwidth}
        \includegraphics[width=\textwidth]{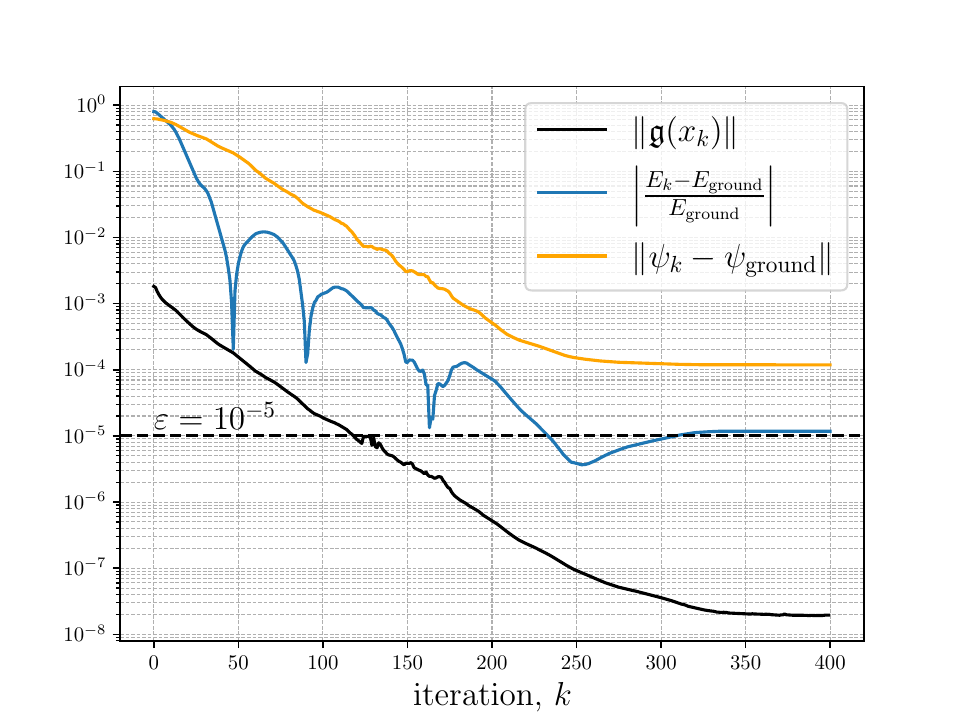}
        \caption{
            Formulation \ref{SCF_formulation02}
        }
    \end{subfigure}
    \hfill
    \begin{subfigure}{0.45\textwidth}
        \includegraphics[width=\textwidth]{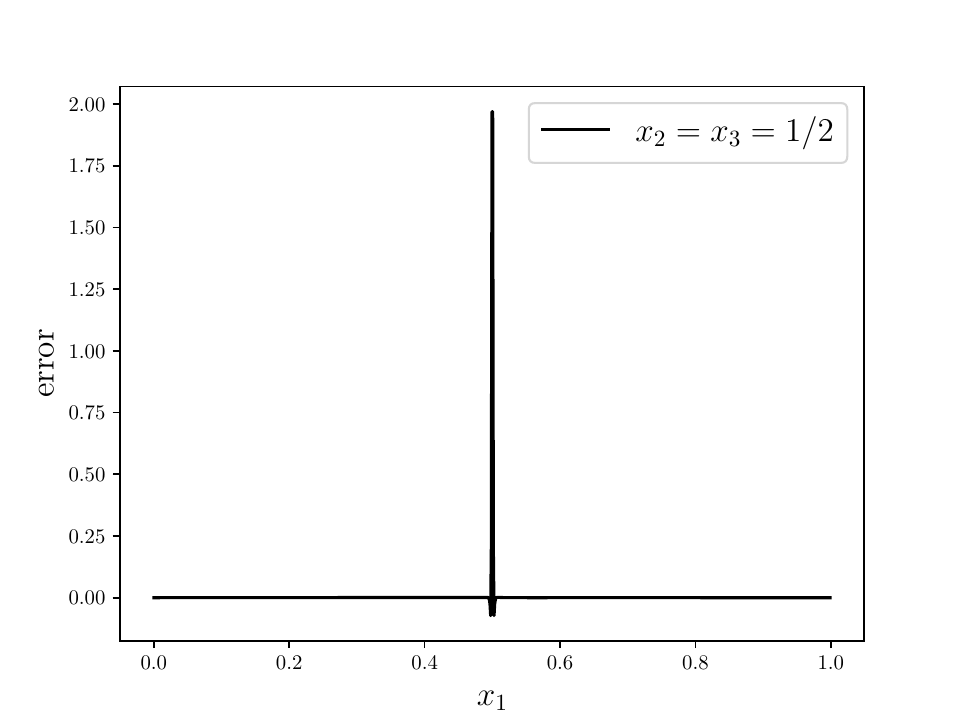}
        \caption{
            Difference between $\psi_{400}$
            associated to \ref{SCF_formulation02}
            and the ground state $\psi_{\text{ground}}$.
        }
    \end{subfigure}
    \hfill
    \caption
    {
        Convergence of DIIS for the hydrogen atom.
        The Laplace operator $\Delta$ is approximated by $\Delta_3(\tau)$
        with $\tau = 10^{-6}$ and the \ac{MRA} order
        $\mathfrak k = 10$ and threshold $\varepsilon = 10^{-5}$.
    }
\label{hydrogen_convergence_figure}
\end{figure}

We consider Equation \eqref{ground_schrodinger} with
\(
    V( \psi, x)
    =
    - 2 L \abs {x - 1/2}^{-1} \psi(x)
    , \
    x \in \mathbb R^3
    ,
\)
where $2L$ is the size of the computational domain in the atomic units.
We set $L = 20$ and restrict ourselves to the unit cube $x \in [0, 1]^3$.
In order to apply the heat approximation for Laplacian,
one has to smear the Coulomb potential $V$ as
\begin{equation}
\label{nuclear_potential_smoothing}
    V( \psi, x)
    \approx
    -
    \frac {2L}{\xi}
    U \left( \frac {\abs {x - 1/2}}{\xi} \right)
    \psi(x)
    , \quad
    U(r)
    =
    \frac {\erf(r)}r
    +
    \frac {e^{-r^2}}{\sqrt \pi}
    ,
\end{equation}
where $\xi$ is a small positive parameter that we set to $\xi = 0.001$.
As an initial guess $\psi_{\text{guess}}$,
we take the function given in \eqref{derivative_test_function_f}
with $\sigma = 1 / (2L)^2 = 6.25 \cdot 10^{-4}$ and normalized.
The Laplace operator $\Delta$
is approximated by the third order expansion $\Delta_3(\tau)$
with $\tau = 10^{-6}$.
The \ac{MRA} parameters are set as $\mathfrak k = 10$ and $\varepsilon = 10^{-5}$.
We initialize $E_{\text{guess}} = \mathfrak G(\psi_{\text{guess}})$
and perform one iteration of \eqref{simple_iteration} to define the starting values
\(
    \psi_0
    ,
    E_0
\)
as in \eqref{starting_values}.
Subsequently, we employ the DIIS accelerator using the formulations
\ref{SCF_formulation00}-\ref{SCF_formulation02}
with $0 \leqslant k - \ell(k) \leqslant 11$.

The ground state, corresponding to
$E = E_{\text{ground}} = - 2L^2 = -800$
and
\[
    \psi(x)
    =
    \psi_{\text{ground}}(x)
    =
    \sqrt{
        \frac {8L^3}{\pi}
    }
    e^{ - 2L \abs {x - 1/2} }
    ,
\]
is used for comparison with the computed values $\psi_k(x)$.
The convergence to the analytical result is demonstrated in Figure \ref{hydrogen_convergence_figure}.
As expected, the largest error occurs at the location of the nucleus.
Nevertheless, the error, with a magnitude of $2$,
is small compared to the ground state peak value, $\psi_{\text{ground}}(1/2, 1/2, 1/2) \approx 142.7$.

A direct discretization of the Laplacian in multiwavelets encounters the same issue as other direct methods,
such as Gauss-type basis representations:
there is little correspondence between the residual norm $\norm{\mathfrak{g}(x_k)}$
and the error in the iteration $\psi_k$.
This lack of correspondence leads to ambiguity in defining the tolerance
required for truncating the iteration scheme.
Consequently, our approach cannot replace the method based on Green's function discretization in multiwavelets.
As noted in \cite{Jensen_Durdek_Bjorgve_Wind_Fla_Frediani2023},
the tolerance $\epsilon$ can be set to match the \ac{MRA} threshold $\varepsilon$,
and the integral formulation converges considerably faster.

%
%
%
%

\subsection{Virtual orbitals}

\begin{figure}[ht!]
    \centering
    \begin{subfigure}{0.49\textwidth}
        \includegraphics[width=\textwidth]{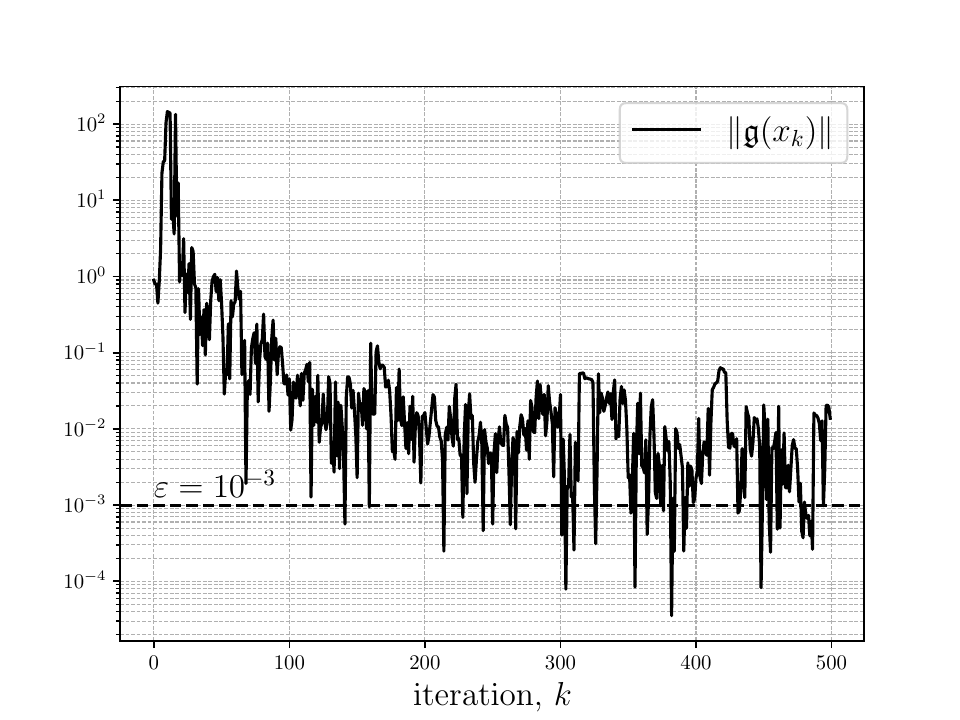}
    \end{subfigure}
    \hfill
    \begin{subfigure}{0.49\textwidth}
        \includegraphics[width=\textwidth]{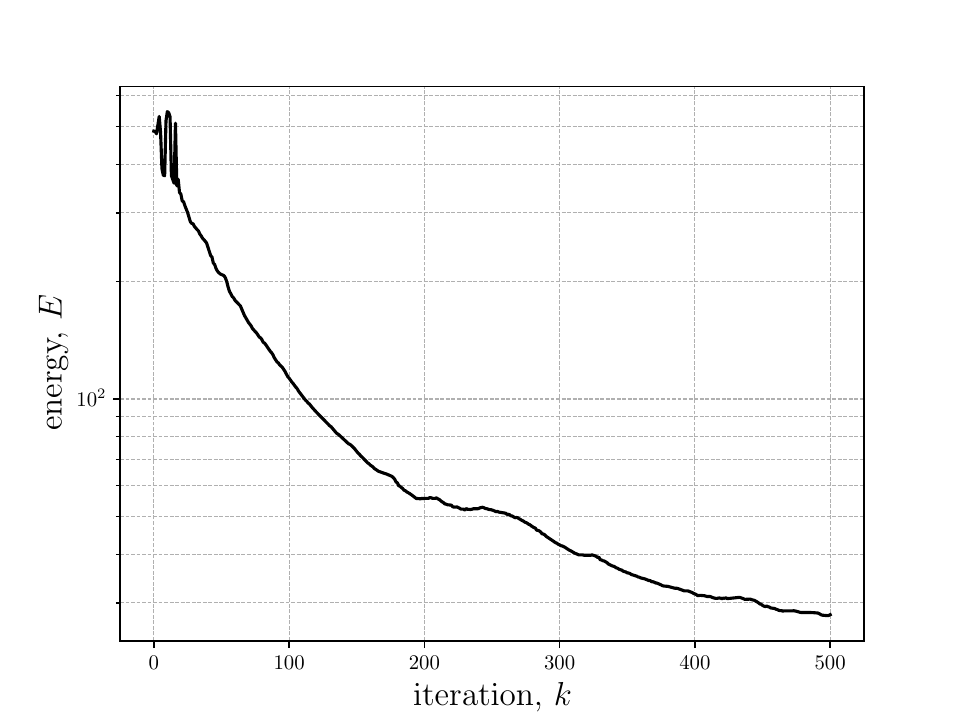}
    \end{subfigure}
    \hfill
    \caption
    {
        DIIS search for the 2s virtual orbital of the helium atom
        associated to Formulation \ref{SCF_formulation00}.
        The Laplace operator $\Delta$ is approximated by $\Delta_3(\tau)$
        with $\tau = 10^{-6}$ and the \ac{MRA} order
        $\mathfrak k = 10$ and threshold $\varepsilon = 10^{-3}$.
    }
\label{helium_convergence_figure}
\end{figure}

We consider closed shell restricted Hartree-Fock equations for $2N$ electrons
in its canonical form in the atomic units
\begin{equation}
\label{general_hartree_fock}
    - \frac 12 \Delta \varphi_i(x)
    +
    V_{\text{nuc}}(x) \varphi_i(x)
    +
    2 \sum_{j = 1}^N
    \int_{\mathbb R^3}
    \frac{\Abs{\varphi_j(y)}^2}{\Abs{x - y}} dy \varphi_i(x)
    -
    \sum_{j = 1}^N
    \int_{\mathbb R^3}
    \frac{\varphi_i(y) \overline{\varphi_j(y)}}{\Abs{x - y}} dy \varphi_j(x)
    =
    \varepsilon_i \varphi_i(x)
\end{equation}
where $x \in \mathbb R^3$ and $i = 1, \ldots, M$ with $M \geqslant N$.
Here $V_{\text{nuc}}$ stands for the nuclei potential.
We are interested in real-valued solutions
$\varepsilon_i \in \mathbb R$, $\varphi_i \in L^2\left( \mathbb R^3 \right)$
with $i = 1, \ldots, M$ satisfying
\begin{equation}
\label{general_normalization}
    \int_{\mathbb R^3}
    \varphi_i(x)
    \varphi_j(x)
    dx
    =
    \delta_{ij}
    \quad
    \text{and}
    \quad
    \varepsilon_1 \leqslant \varepsilon_2 \leqslant \ldots \varepsilon_M
    .
\end{equation}
We refer to the first $N$ functions as occupied (or ground state) spatial orbitals.
The remaining functions, in case $M > N$, are called unoccupied (or virtual) orbitals.
Alternatively, the problem can be approached in two steps: first, solve the system of 
$N$ nonlinear equations; then, address the spectral problem for the resulting Fock operator.

It is implied that $V_{\text{nuc}}(x) \to 0$ for big $|x|$.
More precisely,
$V_{\text{nuc}}(x) \sim -2N / |x|$ as $|x| \to \infty$.
Therefore,
$V_{\text{nuc}}(x)$ together with the Coulomb potential
(the first sum of convolutions in \eqref{general_hartree_fock}),
asymptotically behaves as $C / |x|^2$.
Thus, intuitively, the spectrum of the Hartree-Fock oparator is continuous above zero.
Moreover, the addition of Coulomb and exchange potentials
lifts the limiting discrete hydrogen-type energies into the continuum.
As a result, there are very few virtual orbitals with $\varepsilon_i < 0$.
In practice, direct methods reduce the problem to a matrix eigenvalue problem,
effectively discretizing the continuous spectrum.
Interestingly, the positive energy virtual orbitals,
although dependent on the discretization, may still prove useful for post-Hartree-Fock methods.
Therefore, we address the problem of finding these unoccupied orbitals
using the \ac{MRA} discretization of the heat semigroup representation of the Laplacian introduced above.

We consider helium atom and obtain the ground state using the Green's function approach.
In case of a two electron system we have $N = 1$ and \eqref{general_hartree_fock}
simplifies to
\begin{equation}
    - \frac 12 \Delta \varphi_i(x)
    +
    V_{\text{nuc}}(x) \varphi_i(x)
    +
    4 L
    \int_{\mathbb R^3}
    \frac{\Abs{\varphi_1(y)}^2}{\Abs{x - y}} dy \varphi_i(x)
    -
    2 L
    \int_{\mathbb R^3}
    \frac{\varphi_i(y) \varphi_1(y)}{\Abs{x - y}} dy \varphi_1(x)
    =
    \varepsilon_i \varphi_i(x)
    ,
\end{equation}
where we have also changed the units,
so that the atomic computational domain $[-L, L]^3$ is transformed to $[0, 1]^3$.
The nuclear potential
\(
    V_{\text{nuc}}(x)
    =
    - 4 L \abs {x - 1/2}^{-1}
\)
is smoothed as in \eqref{nuclear_potential_smoothing} with $\xi = 0.001$
and the size of the original domain is set as $L = 20$.
Furthermore, if $\varepsilon_i < 0$ then
there exists a resolvent
$$
    \left( - \Delta - 2 \varepsilon_i \right)^{-1}
    \approx
    H \left( \sqrt{ -2 \varepsilon_i } \right)
    =
    \sum_k c_k\left( \sqrt{ -2 \varepsilon_i } \right) e^{t_k\left( \sqrt{ -2 \varepsilon_i } \right) \Delta}
$$
For $\varepsilon_i \geqslant 0$ we use the direct discretisation described above.
Convolution with $1/\Abs{x}$ is $4\pi H(0)$.

An SCF type system is normally solved iteratively.
It is usually straightforward to choose an initial guess $\varphi_1^0(x)$
for the ground state,
so that the subsequent values $\varepsilon_1^0, \varepsilon_1^1, \ldots < 0$.
Therefore, the first orbital equation
can be rewritten in the integral form
\begin{equation*}
    \varphi_1
    =
    -2 H_x \left( \sqrt{ -2 \varepsilon_1 } \right)
    \left(
        V_{\text{nuc}}(x) \varphi_1(x)
        +
        8 \pi L
        H_y(0) \left( \Abs{\varphi_1(y)}^2 \right) (x) \varphi_1(x)
    \right)
\end{equation*}
that is straightforward to iterate using DIIS, for example.
For the virtual orbitals we will inevitably get
$\varepsilon_i^n \geqslant 0$.
Therefore, the discretization is introduced as
\begin{multline*}
    \varphi_i
    =
    -
    \frac{ T_n(\tau) }{ a_n } \varphi_i(x)
    -
    \frac{2 \tau}{a_n} \varepsilon_i \varphi_i(x)
    \\
    +
    \frac{2 \tau}{a_n}
    \left(
        V_{\text{nuc}}(x) \varphi_i(x)
        +
        16 \pi L
        H_y(0) \left( \Abs{\varphi_1(y)}^2 \right) (x) \varphi_i(x)
        -
        8 \pi L
        H_y(0) \left( \varphi_1(y)\varphi_i(y) \right) (x) \varphi_1(x)
    \right)
\end{multline*}
for $\Delta \approx \Delta_n$ defined by \eqref{recalling_delta}.
We search for a 2s orbital corresponding to $i = 2$.
For $\varphi_2, \varepsilon_2$ we have an SCF type system \eqref{SCF_type_system}
complemented by
\(
    \langle \phi_1 | \phi_2 \rangle = 0
    .
\)
The corresponding DIIS algorithm
for each of the formulations \ref{SCF_formulation01}, \ref{SCF_formulation02} diverges.
In the case of \ref{SCF_formulation00} we have a pattern depicted in Figure \ref{helium_convergence_figure},
suggesting a possibility of defining a virtual orbital.
By picking an orbital with sufficiently small residual, for instance.
However, it is clear that the Fock operator has only one point, namely $\varepsilon_1$,
in the discrete part of spectrum.
Therefore, there is no clear evidence of convergence in Figure \ref{helium_convergence_figure}.

On the other hand,
we have a fixed discretization independent of the iteration $k$,
similar to the basis and grid methods reducing the problem to matrix eigenvalue equations.
Therefore, one may anticipate a discrete spectrum with a uniquely defined point $(\varphi_2, \varepsilon_2)$
for our discrete version of the Fock operator.
The author believes that the reason for this inconsistency might come from the way we eliminate the noise
in multiwavelet calculations in practice.
Indeed, expanding convolution operators, including Laplacian $\Delta$,
in heat semigroup as in \eqref{resolvent_approximation} we bring some numerical noise
into our iterative scheme.
This noise is below the \ac{MRA} threshold $\varepsilon$,
and its main effect is an increase of memory allocation for each new iteration.
Thus, this noise can be filtered \cite{Bjorgve_2024_vampyr}.
It is worth to notice,
that when a kernel is treated exactly, as for example in the time evolution treatment
in \cite{Dinvay2024}, the filtering is not needed.
Under this necessary filtering the discretization is not completely fixed
and this might cause oscillatory behavior in Figure \ref{helium_convergence_figure}.

\section{Free-particle propagator}
\label{Free_particle_propagator_section}
\setcounter{equation}{0}

This section is devoted to the numerical discretisation of the exponential operator
\(
    \exp \left( i t \Delta \right)
    ,
\)
where $t > 0$ is a small parameter,
using the heat representation $\Delta \approx \Delta_n(\tau)$.
More precisely, we aim to obtain an expansion of the form
\begin{equation}
\label{propagator_approximation}
    \exp \left( i t \Delta \right)
    \approx
    \sum_{j} c_j e^{t_j \Delta}
\end{equation}
in a manner similar to \eqref{resolvent_approximation}.
The representation \eqref{propagator_approximation} is not necessarily the most
efficient way of utilizing the discretization $\Delta \approx \Delta_n(\tau)$ in
dynamical simulations.
In the future, we plan to focus more on performance and applied problems.
For such purposes, the use of Krylov subspace methods should, at least intuitively,
be more suitable than a Taylor series expansion of the exponential.
Here, however, we aim to express the free-particle propagator
using the minimal possible number of heat operators in \eqref{propagator_approximation}.

Naively, one can set $\tau = t$
and expand
\begin{equation}
\label{naive_expansion}
    e^{i t \Delta}
    =
    \sum_{m = 0}^N
    \frac
    { ( it \Delta_{N - m + 1}(t) )^m }
    { m! }
    +
    O \left( t^{N + 1} \right)
    .
\end{equation}
However, utilizing \eqref{naive_expansion}
may lead to unstable numerical algorithms.
Indeed,
in the second-order case $N = 2$, the approximation of the Schr\"odinger semigroup
\eqref{naive_expansion}
simplifies to
\begin{equation*}
    e^{i t \Delta}
    =
    \frac 12 - \frac 32i
    +
    (1 + 2i) e^{t \Delta}
    -
    \frac 12 (1 + i) e^{2t \Delta}
    +
    O \left( t^3 \right)
    .
\end{equation*}
This is a Fourier multiplier, as are all its approximations.
The symbol of $e^{t \Delta}$ takes all values in $(0, 1]$
in frequency domain,
and thus the norm of the above approximation is
\begin{equation*}
    \norm
    {
        \frac 12 - \frac 32i
        +
        (1 + 2i) e^{t \Delta}
        -
        \frac 12 (1 + i) e^{2t \Delta}
    }
    =
    \sup_{x \in [0, 1]}
    \left|
        \frac 12 - \frac 32i
        +
        (1 + 2i) x
        -
        \frac 12 (1 + i) x^2
    \right|
    =
    \sqrt{\frac 52}
    .
\end{equation*}
Since this norm is greater than one, it can lead to numerical instability in practical calculations.

For sufficiently small $x$ (depending on $M$),
the Taylor polynomials $P_M(\cos, x)$ and $P_M(\sin, x)$,
where $M + 1$ represents the number of terms in their expansions, do not exceed $|\cos x|$ and $|\sin x|$,
respectively, provided $M$ is odd.
Consequently, for the Taylor polynomial
\begin{equation}
    P_{2M + 1}(x)
    =
    P_M(\cos, x)
    +
    i P_M(\sin, x)
    =
    \sum_{m = 0}^M
    \frac{ (-1)^m }{(2m)!} x^{2m}
    +
    i
    \sum_{m = 0}^M
    \frac{ (-1)^m }{(2m + 1)!} x^{2m + 1}
\end{equation}
of the exponential $e^{ix}$ with odd positive integer $M$,
we have $|P_{2M + 1}(x)| \leqslant 1$ provided $|x| \leqslant \sqrt 2$.
Where we took into account that the first positive root of $P_1(\cos, x)$
can serve as an obvious uniform bound for $|x|$ respecting the inequality.
Therefore,
substituting instead of $x$ a Fourier multiplier with the norm bounded by $\sqrt 2$
we obtain a new pseudo-differential operator with the norm bounded by 1.
On the other hand,
$P_{2M + 1}(0) = 1$
and we know that $x = 0$ in the limit $\xi \to \infty$
for every Laplacian approximation.
Then the free-particle semigroup $\exp \left( it \Delta \right)$ can be approximated by
$P_{2M + 1}(t \Delta_{2M + 1}(\tau))$ with $M = 1, 3, \ldots$,
each having the norm equaled 1,
provided $\norm{ t \Delta_{2M + 1}(\tau) } \leqslant \sqrt 2$.

Setting $\tau = t / a$, we finally arrive at an expression similar to \eqref{propagator_approximation}
reading
\begin{equation}
\label{M_propagator_approximation}
    e^{i t \Delta}
    =
    P_{2M + 1}
    \left(
        a \tau \Delta_{2M + 1}(\tau)
    \right)
    +
    O \left( t^{2M + 2} \right)
    ,
\end{equation}
where $\tau \Delta_{2M + 1}(\tau)$ is a polynomial with respect to $e^{\tau \Delta}$.
For example, when $M = 1$, corresponding to the third order approximation,
the norm of the heat semigroup representation in \eqref{M_propagator_approximation} equals 1,
provided $a = 6 \sqrt 2 / 11$, see Table \ref{laplace_degree_norm_table}.
The bound $a$ is not optimal and can be increased up to
$a = 6 \sqrt 2 / 11 + 5/29$ without spoiling the isometry of the approximation.
For $M = 3$, the representation \eqref{M_propagator_approximation}
contains an impractically large number of terms, making it unsuitable for practical applications.

A question arises: can the number of heat operators in the final expansion be reduced
by considering $\Delta_n$ of different orders $n$ in \eqref{M_propagator_approximation}, instead?
In other words, we aim to find an expansion of the form
\begin{equation}
\label{M_propagator_approximation_optimized}
    e^{i t \Delta}
    \approx
    \sum_{m = 0}^N
    \frac
    { ( ia \tau \Delta_{N - m + 1} (\tau) )^m }
    { m! }
    =:
    \mathcal P(a, x)
    , \quad
    x = e^{\tau \Delta}
    , \quad
    N = 2M + 1
    , \quad
    M = 1, 3, \ldots.
\end{equation}
and a number $a \sim 1$ depending on $M$ such that
$|\mathcal P(a, x)|^2 \leqslant 1$ for all $x \in [0, 1]$.
It turns out that such optimization is not possible for $M = 1$,
but it is achievable for $M = 3$ with $a = 1/2$.
This gives the final optimal approximation \eqref{M_propagator_approximation_optimized}
for the order $N = 2M + 1 = 7$,
involving 36 heat operators in \eqref{propagator_approximation} excluding identity.
Surprisingly,
the higher order exponential operator approximation ($N = 7$)
performs worse than the lower-order approximation ($N = 3$)
in terms of both precision and memory usage.
Therefore, it is omitted in the simulations below.

Finally, we present some practical representations for the approximation of $e^{it \Delta}$,
each expressed in the form \eqref{propagator_approximation}.
The time $t$ and the approximation parameter $\tau$ are related by $t = a \tau$.
\begin{enumerate}
    \item
    Second order representation:
    \begin{equation}
    \label{final_propagator_approximation_2}
        e^{it \Delta}
        =
        1
        -
        \frac 12
        (a \tau \Delta_2 (\tau))^2
        +
        i a
        \left(
            \tau \Delta_2 (\tau)
            -
            \frac 16
            (\tau \Delta_1 (\tau))^3
        \right)
        +
        O \left( t^3 \right)
        , \quad
        a = \frac 35
        .
    \end{equation}
    \item
    Taylor expansion:
    \begin{equation}
    \label{final_propagator_approximation_34}
        e^{i t \Delta}
        =
        \sum_{m = 0}^N
        \frac
        { ( ia \tau \Delta_N (\tau) )^m }
        { m! }
        +
        O \left( t^{N + 1} \right)
        , \quad
        \left[
        \begin{aligned}
            &
            \,
            N = 3
            , \quad
            a = \frac {6 \sqrt 2}{11} + \frac 5{29}
            ,
            \\
            &
            \,
            N = 4
            , \quad
            a = \frac{25}{21}
            .
        \end{aligned}
        \right.
    \end{equation}
    \item
    Sixth order representation:
    \begin{equation}
    \label{final_propagator_approximation_6}
        e^{i t \Delta}
        =
        \sum_{m = 0}^4
        \frac
        { ( ia \tau \Delta_6 (\tau) )^m }
        { m! }
        +
        \sum_{m = 5}^7
        \frac
        { ( ia \tau \Delta_3 (\tau) )^m }
        { m! }
        +
        O \left( t^7 \right)
        , \quad
        a = \frac{100}{229}
        .
    \end{equation}
\end{enumerate}

\subsection{Harmonic potential}

\begin{figure}[ht!]
    \centering
    {
	\includegraphics[width=0.7\textwidth]
        {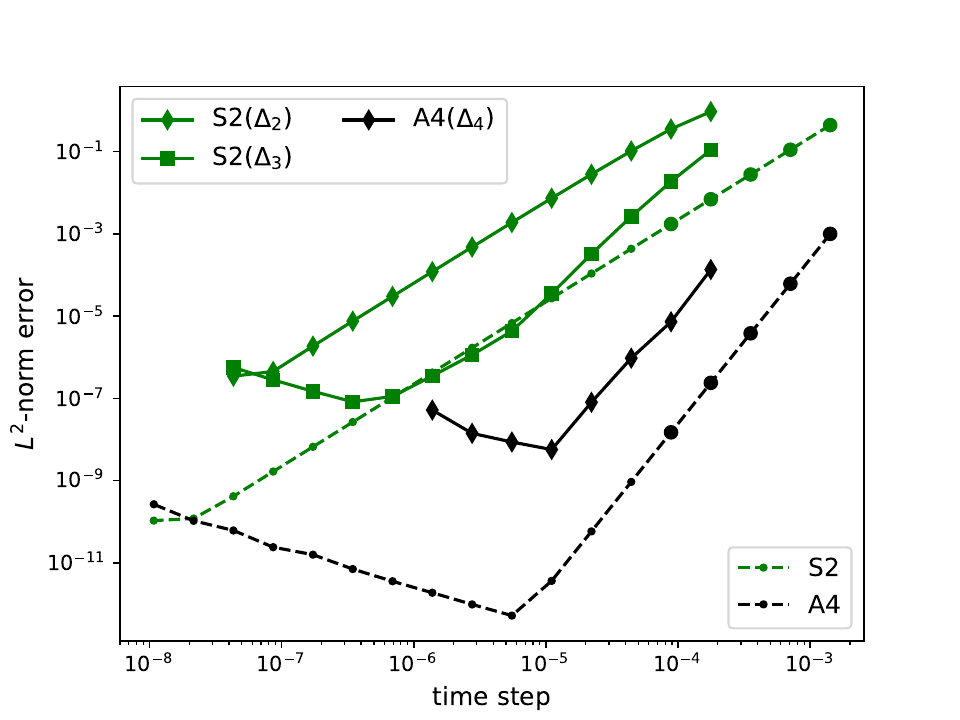}
    }
    \caption
    {
	Convergence of time evolution in the harmonic potential.
        The reference is formed by \eqref{S2_integrator}, \eqref{Chin_Chen_A} utilizing
        FFT-based schemes (dashed lines) and
        MRA-based schemes relying on the contour deformation
        of the ossiclatory integrals \cite{Dinvay2024} (large dots).
        The schemes $\text{S2}(\Delta_2)$ and $\text{S2}(\Delta_3)$, $\text{A4}(\Delta_4)$
        are based on approximations \eqref{final_propagator_approximation_2} and
        \eqref{final_propagator_approximation_34}, respectively.
    }
\label{harmonic_convergence_figure}
\end{figure}

We test the proposed expansions for the free particle propagator
by simulating the one dimensional Schrödinger equation \eqref{time_schrodinger}
reading
\begin{equation*}
    i \partial_t \psi
    =
    - \frac 12 \partial_x^2 \psi
    + V(x) \psi
    , \quad
    x \in \mathbb R
\end{equation*}
with the harmonic potential
\begin{equation*}
    V(x)
    =
    V_0 \left( x - \frac 12 \right) ^2
    .
\end{equation*}
It is complemented by the initial condition
$\psi(x, 0) = \psi_0(x) = f(x)$ having the Gaussian form
\eqref{derivative_test_function_f}.
It is a matter of common kwnoledge that in the harmonic potential
the density
$
    \left| \psi(t) \right|^2
$
oscillates with the period
$
    t_{\text{per}} = \pi \sqrt{2 / V_0}
    ,
$
more precisely,
$
    \psi \left( t_{\text{per}} \right) = - \psi_0
    .
$
We perform simulations on the unit space interval $[0, 1]$,
setting the parameters
$V_0 = 98304$ for the potential,
$x_0 = 0.375$ and $\sigma = 0.0025$
for the initial function $\psi_0 = f(x)$ given in \eqref{derivative_test_function_f}.
The numerical simulations are carried out with time steps of $t_{\text{per}} \cdot 2^{-m} / 10$.
We use symplectic integrators of the second order
\begin{equation}
\label{S2_integrator}
\tag{S2}
    \exp \left(  i \frac t2  \Delta  - it V \right)
    =
    \exp \left( - i \frac t2 V \right)
    \exp \left(  i \frac t2  \Delta \right)
    \exp \left( - i \frac t2 V \right)
    +
    O \left( t^3 \right)
    .
\end{equation}
and of the fourth order
\begin{multline}
\label{Chin_Chen_A}
\tag{A4}
    \exp \left(  i \frac t2  \Delta  - it V \right)
    =
    \exp \left(  -i \frac t6  V \right)
    \exp \left(  i  \frac t4  \Delta \right)
    \exp
    \left(
        -i \frac {2t}3
        \left(             
            V
            -
            \frac{t^2}{48}
            ( \partial_x V )^2
        \right)
    \right)
    \\
    \cdot
    \exp \left(  i  \frac t4  \Delta \right)
    \exp \left(  -i \frac t6  V \right)
    +
    O \left( t^5 \right)
    ,
\end{multline}
the latter originated in \cite{Suzuki1995}.
Combining \eqref{S2_integrator} with \eqref{final_propagator_approximation_2}
one obtains a second order scheme, denoted by $\text{S2}(\Delta_2)$.
Combining \eqref{S2_integrator} with the third order expansion \eqref{final_propagator_approximation_34},
corresponding to $N = 3$,
one obtains a second order scheme, denoted by $\text{S2}(\Delta_3)$.
Finally, \eqref{Chin_Chen_A} together with the fourth order expansion \eqref{final_propagator_approximation_34},
corresponding to $N = 4$,
gives a fourth order scheme which we denote as $\text{A4}(\Delta_4)$.
Note that \eqref{Chin_Chen_A} could recursively be transformed into a sixth order scheme,
as detailed in \cite{Dinvay2024}.
Then using it together with expansion \eqref{final_propagator_approximation_6}
would give rise to a sixth order scheme.
However,
our simulations indicated that the resulting accuracies
were comparable to those achieved by the fourth-order scheme $\text{A4}(\Delta_4)$,
while the computational complexity and resource requirements were significantly higher.
Therefore, the sixth-order numerical results are omitted.

The accuracy of the obtained numerical solution $\psi$ for each of these schemes
is estimated by the $L^2$-difference
\begin{equation}
\label{error}
    \text{error}
    =
    \norm{ \psi(t_{\text{per}}) + \psi_0 }_{L^2(0, 1)}
    .
\end{equation}
The \ac{MRA} tolerance needed for adaptive resolution is set to $\varepsilon = 10^{-10}$.
In the schemes $\text{S2}(\Delta_2)$, $\text{S2}(\Delta_3)$ and $\text{A4}(\Delta_4)$
we take the MRA order $\mathfrak k = 10$.
The performance is compared with the 
Fourier spectral method calculations using $1024$ grid points inside of the unit interval $[0, 1]$.
The results are provided
in Figure \ref{harmonic_convergence_figure},
which also includes the corresponding multiwavelet calculations from \cite{Dinvay2024}.
These calculations exploit the evaluation of highly oscillatory integrals
and are limited to large time steps and a higher MRA order, set to $\mathfrak k = 18$.

The $\psi$-function dynamics in the harmonic potential well is smooth
and it can be viewed as periodic with very high precision.
Consequently, the spectral method yields curves
in Figure \ref{harmonic_convergence_figure}
that can be considered ideal.
The \ac{MRA} calculations from \cite{Dinvay2024}
align with these curves, as the exponential is treated exactly in the frequency domain.
In contrast, the curves corresponding to the schemes
$\text{S2}(\Delta_2)$, $\text{S2}(\Delta_3)$ and $\text{A4}(\Delta_4)$
are slightly elevated above the ideal ones due to the approximation of the exponential.
However, the new approach is less constrained by the time step size and the \ac{MRA} order,
offering greater flexibility.

\section{Conclusion}
\setcounter{equation}{0}

We introduced an alternative \ac{MRA} representation of the Laplacian,
with the key advantage of supporting iterative application.
This representation is particularly well-suited for peaked functions, which naturally arise in many-electron systems.
As a result, it has the potential to enhance the robustness of existing multiwavelet algorithms in quantum chemistry.
Moreover,
it may extend the use of \ac{MRA} to those areas,
where the Green's function formulation is not applicable.
Additionally, the approach enables direct identification of virtual orbitals within the multiwavelet framework.

The new Laplacian discretization also extends the applicability
of the previously introduced time evolution operator \cite{Dinvay2024, Dinvay_Zabelina_Frediani2024}.
It achieves this by removing the \ac{MRA} order restriction and alleviating constraints on the time step size.
As a matter of fact,
expanding the use of multiwavelets in dynamical problems was the primary motivation for this work.



\vskip 0.05in
\noindent
{\bf Acknowledgments.}
{
    The author is grateful to
    R.Vikhamar-Sandberg, Ch. Tantardini and L. Frediani
    for numerous helpful discussions.
    I acknowledge support from the Research Council of Norway through
    its Centres of Excellence scheme (Hylleraas centre, 262695)
    and the support from
    the FRIPRO grant ReMRChem (324590) from the Research Council of Norway
    hold by L. Frediani.
}

\bibliographystyle{acm}
\bibliography{bibliography}

\end{document}